\documentclass[10pt]{article} %
\usepackage{geometry}
\usepackage{graphicx}
\usepackage{epsfig}
\usepackage{amsmath}
\usepackage{csquotes}
\usepackage{ar}
\usepackage{amsfonts}
\usepackage{amssymb}
\usepackage{graphicx}
\usepackage{wrapfig}
\usepackage{sidecap}
\usepackage{subfigure}
\usepackage{multicol}
\usepackage{multirow}
\usepackage{rotating}
\usepackage{tikz}

\usepackage[colorlinks=true,
			linkcolor=blue,
			linktoc=page,
			citecolor=blue,
			urlcolor=blue,
			filecolor=blue,
			breaklinks=true]{hyperref}

\usepackage{float}
\usepackage{csquotes}
\usepackage{cleveref}
\usepackage{comment}
\usepackage{booktabs}
\usepackage{soul}

\usepackage{mathrsfs}

\newcommand{\dt}[1]{\mathrm{d}#1}
\newcommand{\overbar}[1]{\mkern 1.5mu\overline{\mkern-1.5mu#1\mkern-1.5mu}\mkern 1.5mu}

\usepackage[authoryear]{natbib}
\bibpunct{(}{)}{;}{a}{,}{,}
\title{The influence of freestream turbulence on the near-field growth of a turbulent cylinder wake: turbulent entrainment and wake meandering}
\author{K.S. Kankanwadi and O.R.H. Buxton\\ \small{Department of Aeronautics, Imperial College London, U.K.}} 
\date{}

\begin{document}
\maketitle

\section*{Abstract}
The effect of freestream turbulence on the spreading of the near wake of a circular cylinder ($< 11$ cylinder diameters from the rear face of the cylinder) is investigated through particle image velocimetry (PIV) data.
Different ``flavours'' of freestream turbulence, in which the turbulence intensity and integral length scale are independently varied, are subjected to the cylinder.
The time-averaged spreading of the near wake is de-coupled into the growth through entrainment of background fluid and the envelope of the spatial extent of the instantaneous wake due to wake ``meandering'', induced by the presence of large-scale vortical coherent structures, i.e. the von K\'{a}rm\'{a}n vortex street.
Unlike for the far-field of a turbulent wake, examined by \citet{Kankanwadi2020}, it is shown that freestream turbulence enhances the entrainment rate into the wake in comparison to a non-turbulent background.
Further, both the turbulence intensity and the integral length scale of the background turbulence are important in this regard, further contrasting to the far wake where only the turbulence intensity is important.
Additionally, wake meandering is enhanced by the presence of background turbulence, and here the integral length scale is the dominant parameter.
Combining these findings yields the oft-reported result that background turbulence enhances the time-averaged near-wake growth rate.
The influence of wake meandering is isolated by conducting similar experiments in which a splitter plate is mounted to the rear face of the cylinder, thereby eliminating the von K\'{a}rm\'{a}n vortex street.
These results show that when large-scale vortices aren't present within the turbulent wake then freestream turbulence actually suppresses the entrainment rate, relative to a non-turbulent background, in results that mirror \citet{Kankanwadi2020}.
We therefore postulate that freestream turbulence has the effect to enhance large-scale entrainment via engulfment, but suppress entrainment via small-scale nibbling.
Finally, we observe that the presence of freestream turbulence occasionally leads to a transient elimination of vortex shedding, an effect that is bound to have consequences on the instantaneous entrainment rate as outlined above. 

\section{Introduction}

Predicting the spreading rate of turbulent wakes is important for a number of of applications, including designing the layout of wind farms for maximum efficiency \citep{Meyers2012} and predicting the influence of tall buildings in generating gusts \citep{Xie2008} and dispersing pollution \citep{Aristodemou2018} in urban environments.
Much work has been done on studying canonical turbulent wakes produced by circular cylinders in a uniform flow, see for example the review paper of \citet{Williamson1996}, however in both scenarios listed above the turbulent wake is in fact exposed to a turbulent background.
The near-field development of a cylinder wake in the presence of background turbulence has been briefly examined in the literature. 
\cite{Bearman1983} examined the effects of free-stream turbulence on the mean flow over bluff bodies in the near field. 
They summarised the main effects to be accelerated transition of the wake to turbulence, and enhanced mixing and entrainment. 
Furthermore, they highlighted the importance of coherent structures on initial shear layer development and speculated on the effect that free-stream turbulence will have on them. 
Lastly, they outlined the fact that the mechanisms by which free-stream turbulence affects spreading rate were not yet well understood.

\cite{eames2011growth} addressed the problem more recently. 
They developed a theoretical model of wake spreading downstream of a cylinder in ambient turbulence. 
They report that wake spreading gets affected by freestream turbulence when the velocity deficit has decayed sufficiently that it becomes comparable to the background turbulence intensity. 
Their model predicts that when the mean wake width is smaller than the integral length scale of the freestream turbulence, the wake grows linearly with streamwise distance ($\overline \delta \sim x$). 
Contrastingly, when the wake width is larger than the integral length scale the wake is predicted to grow diffusively ($\overline \delta \sim x^{1/2}$). 
The authors also conducted a laboratory run of flow past a cylinder in intense ambient turbulence (18\%) that consisted of a large integral length scale ($9.5d$ where $d$ is the cylinder diameter). 
They reported that the velocity deficit showed good agreement with their model.
However, the model was only developed for instances where either very weak or very strong turbulence intensity was subjected to the cylinder. 
It does not aim to predict instances where both, internal and external regions of turbulence are significant to the spreading process. 
The authors conclude by stating that further research is required to understand the spreading of a plane wake under the effect of various levels of free-stream turbulence.

Turbulent wakes grow through the process of turbulent entrainment.
This encompasses the physical processes that lead to the transport, and mixing of background fluid across the well-defined interface that separates the turbulent wake from its background. 
Entrainment has been extensively studied in the special context of a turbulent flow expanding into a non-turbulent background where this bounding interface demarcating the primary turbulent flow from the background is known as the turbulent/non-turbulent interface (TNTI).
Particular emphasis has been dedicated to understanding entrainment in the ``fully-developed'' region of turbulent shear flows exposed to a non-turbulent background, i.e. many cylinder/jet diameters downstream: see for example the review of \citet{DaSilva2014}.
In the far-field region entrainment is primarily driven by the process of ``nibbling'', whereby the vorticity within the wake is transmitted to the irrotational background fluid via viscous diffusion, something that was first postulated by \citet{Corrsin1955} and subsequently experimentally verified by \citet{Holzner2007}.
Since diffusion is a viscous phenomenon then entrainment (as characterised by the entrainment velocity) is typically assumed to occur at the smallest velocity scales present within the flow.
Contrastingly, in the near-field of turbulent shear flows (e.g. jets) entrainment is observed to be largely driven by large-scale ``engulfment'' of background fluid \citep{Yule1978}.

The presence of background turbulence complicates the process of entrainment.
\citet{Kankanwadi2020} studied entrainment into the far field of a turbulent wake produced by a circular cylinder (40 diameters downstream) and verified the existence of a turbulent/turbulent interface (TTI) for a broad range of ``flavours'' of background turbulence, characterised by the turbulence intensity $TI$ and integral length scale $L$.
They additionally showed that the presence of freestream turbulence further contorts the TTI (in comparison to the TNTI) with this contortion increasing as the turbulence intensity of the background turbulence increases.
Whilst this enhanced contortion thereby increases the surface area of the interface over which diffusion can take place they reported that background turbulence actually has the effect of attenuating/suppressing the rate of entrainment into the wake in comparison to a non-turbulent background.
This was attributed to the increased prevalence of intermittent, but powerful detrainment events, i.e. a transport of fluid from the wake to the background, as background turbulence intensity was increased.
\citet{Kankanwadi2022} subsequently showed that the physics of the TTI were fundamentally different to those of the TNTI since the inertial vorticity stretching term $\omega_i s_{ij} \omega_j$, where $\omega_i$ is the vorticity vector and $s_{ij}$ is the fluctuating rate-of-strain tensor, dominates viscous diffusion.
$\omega_i s_{ij} \omega_j$, like all velocity-gradient quantities, is an intermittent quantity, perhaps explaining the importance of intermittency in determining the entrainment rate between two adjacent streams of turbulence.

In part due to the complexity of the flow physics of turbulent wakes exposed to background turbulence, empiricism is often used to model the spreading of wind-turbine wakes; an important application for the prediction of turbulent wake spreading as highlighted earlier.
An example of such an empirical model is that reported by \cite{Niayifar2016}. 
They adapt the model laid out by \cite{Bastankhah2014}, to incorporate the effects of background turbulence intensity. 
The proposed empirical model, obtained from large-eddy simulations, explicitly assumes a linear growth rate of the wake with downstream distance, i.e. $\overline \delta \sim x$ (as do most existing empirical wind-turbine models), with the linear growth rate increasing with the intensity of the background turbulence.
Note that this is in contrast to the results of \citet{Kankanwadi2020} that show a reduction in entrainment rate with background turbulence intensity, which should correspond to a reduced wake spreading rate, in the \ul{far field} of a cylinder wake.
For practical reasons these models are focused on predicting the time-averaged spreading rate of the near wake, i.e. $\overline \delta(x)$ is a statistical measure of the mean wake width, substantially closer than $40d$ to the wake generator.
This mean wake width can of course be defined in several ways, for example when making an assumption that a turbulent wake has a Gaussian mean velocity profile \citep[e.g.][]{Bastankhah2014} then the variance of the underlying Gaussian function $\sigma(x)$ can be used.

Underpinning these wake models is the dubious assumption that the mean velocity profile of the wake is self-similar, e.g. can be described with a Gaussian, from very close to the wake generator.
Whilst turbulent wakes do eventually reach a state of self-similarity sufficiently far downstream making a Gaussian representation of the mean velocity profile an accurate one, it is a known limitation of these various wake models that the near-wake region is not well described as being self-similar.
In fact, for a truly self-similar flow (i.e. one in which the turbulence has ``lost its memory'' of the initial conditions) the spreading rate of an axisymmetric turbulent wake can be predicted from the self-similar form of the RANS/turbulent kinetic energy (TKE, $k$) equations with a closure for the dissipation ($\varepsilon$) term of the TKE equation \citep{George1989}.
If a classical closure for the dissipation rate is used, i.e. $\varepsilon = C_\varepsilon k^{3/2}/L$ where $L$ is the integral turbulent length scale and $C_\varepsilon$ is a constant, then \citet{George1989} shows that $\overline \delta \sim x^{1/3}$.
More recently, a new dissipation scaling has been observed for a broad variety of turbulent flows in which the dissipation rate is not in equilibrium with the inter-scale flux of the mean cascade of TKE in which case $\varepsilon = C_\varepsilon k^{3/2} / L$ with $C_\varepsilon \sim Re_G^m / Re_{\mathcal L}^n$ where $Re_G$ is a global (flow) Reynolds number and $Re_{\mathcal L}$ is a local (turbulent) Reynolds number \citep{Vassilicos2015}.
If this dissipation scaling is used for closure in the TKE/RANS equations for an axisymmetric turbulent wake, with the most commonly reported values from literature of $m \approx n \approx 1$, then an alternative scaling $\overline \delta \sim x^{1/2}$ is obtained.

Regardless of which dissipation closure is used it can be seen that neither $\overline \delta \sim x^{1/3}$ nor $\overline \delta \sim x^{1/2}$ match the oft-reported linear spreading, i.e. $\overline \delta \sim x$, of wind-turbine wakes, at least in the region $x \lesssim 15D$ where $D$ is the turbine diameter \citep{Bastankhah2014}.
Whilst entrainment is known to be affected by the presence of freestream turbulence \citep{Kankanwadi2020} there is an additional complexity to the near wake, that is not encountered in the far wake: the presence of energetic coherent structures.
For a cylinder wake, these manifest themselves as the von K\'{a}rm\'{a}n vortex street.
The presence of these large-scale vortices makes the near wake extremely dynamic, with an oscillation or ``meandering'' of the extent of the wake in the transverse direction.
Whilst, the geometry of a wind turbine is significantly different to a cylinder, the near-wake region of a wind-turbine wake nevertheless contains many energetic coherent structures, such as the helicoidal tip vortex \citep{Vermeer2003}.
In either case, the time-averaged wake width in the near field is affected by the envelope of the spatial extent of these large-scale structures.
This is driven by both the entrainment of background fluid into the near wake, and the amplitude of the meandering of the outermost extent of the wake.
The scaling $\overline \delta \sim x^\alpha$ is thus affected by both the meandering of, and entrainment into the near wake, which to the authors' knowledge has not yet been fully de-coupled.
Accordingly, even in the near field of a cylinder wake, a conclusive understanding of the effects of free-stream turbulence on the wake's growth is still lacking. 
The aim of this paper is therefore to address these concerns with a parametric study exploring the near-field development of a cylinder wake exposed to freestream turbulence.
Two separate experimental campaigns are reported: one in which a conventional circular cylinder is exposed to different ``flavours'' of freestream turbulence, and one in which a splitter plate is attached to the rear face of the cylinder in  order to try and isolate the effects of the large-scale coherent von K\'{a}rm\'{a}n vortices on the near-wake development. 

\section{Methodology}

\subsection{Experimental methodology}

\begin{figure}
	\centering
	\includegraphics[width = 0.5\textwidth]{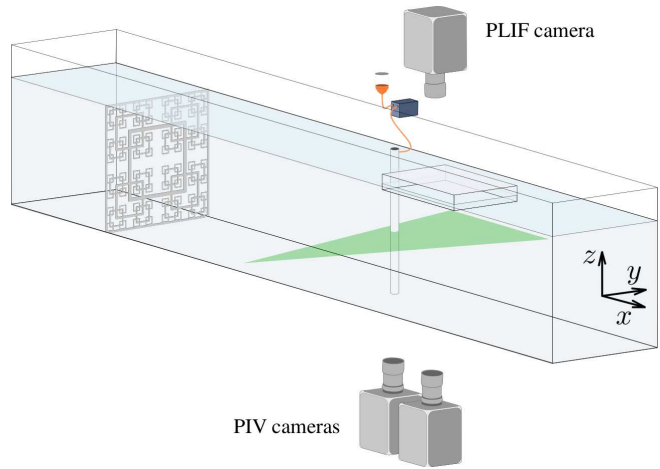}
	\caption{Schematic for the simultaneous PIV/PLIF experiments. Note that the PLIF camera images the flow through a perspex sheet so as to avoid distortion due to the free surface of the water flume. A representative coordinate system is displayed in the figure for convenience; the true origin of the coordinate system is on the rear face of the cylinder at its centre-line.}
	\label{fig:expsetup}
\end{figure}

The experimental methodology consisted of simultaneous particle image velocimetry (PIV) and planar laser induced fluorescence (PLIF) experiments which were conducted in a water flume with a working section length of 9 m and a cross-section of $600 \times 600$ mm$^2$, located at Imperial College London.
A schematic for the experimental set up is shown in figure \ref{fig:expsetup}.
A Reynolds number, based on the diameter $d = $ 30 mm of the circular cylinder, $ Re_d $, of approximately 12,000 was achieved hence placing the flow around the cylinder in the shear-layer transition regime \citep{Williamson1996}. 
Note that the presence of the cylinder alone, resulted in a blockage ratio of 5\%, thereby suggesting that wall effects may be considered negligible. 

\begin{figure}
	\centering
	\includegraphics[width = 0.7\textwidth]{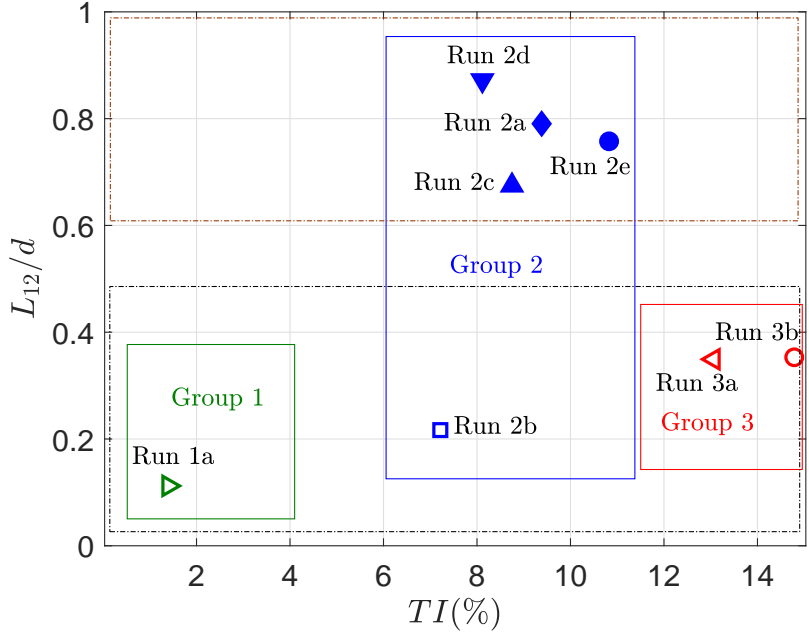}
	\caption{Experimental envelope for the near-wake study. The length scales in this plot have been normalised with the cylinder diameter $d = 30$ mm. The different ``flavours'' of freestream turbulence are divided into three groups based on the turbulence intensity, similarly to \citet{Kankanwadi2020,Kankanwadi2022}. Additionally, two new groupings have been proposed (dot-dashed boxes). These classify ``large'' and ``small'' length scale runs.}
	\label{fig: Near wake exp envelope}
\end{figure}

In this investigation, different ``flavours'' of incoming free-stream turbulence were produced by turbulence-generating grids placed upstream of the circular cylinder. 
The same 4 perspex grids, a combination of two regular and two space-filling square fractal grids, were used as in the studies of \citet{Kankanwadi2020,Kankanwadi2022}.
The philosophy behind the design of these grids was to encourage the $ \{L,TI\} $ parameter space of the incident freestream turbulence to be as widely investigated as possible, in order to truly test the sensitivities of the near-wake physics to both length scale as well as turbulence intensity. 
The flow downstream of all four grids was fully characterised and therefore the cylinder could be strategically placed at varying downstream distances so as to independently vary a single parameter in the background turbulence whilst the other remained constant. 
The experimental envelope illustrating the $ \{L_{12},TI\} $ parameter space of the different ``flavours'' of freestream turbulence is presented in figure \ref{fig: Near wake exp envelope}.
The turbulent length scale $L_{12}$ is computed in the same fashion as in \citet{Kankanwadi2020,Kankanwadi2022}, i.e. $L_{12} = \int_0^{\hat y} R_{12} \dt y$, where $R_{12}$ is the normalised auto-correlation function of the streamwise velocity fluctuations $u'$ in the $y$-direction and $\hat y$ is the location of the first zero crossing of $R_{12}$.
$L_{12}$ is normalised by the cylinder diameter $d$ and $TI$ is averaged across the field of view.
In order to mount these grids in the flume they were encased in a frame and the blockage  of the grids as well as the encasing frame varied in the range of $32\% < \sigma < 35\%$.

\begin{table}
	\centering
	\caption{Experimental parameters used in the experiments. Note that $ f_{aq} $, $ \dt t $ and $ N_{aq} $, represent the acquisition frequency, the time separation between two frames of a single PIV burst and the total number of double frames captured.}
	\label{tb: exp parameters}
	\begin{tabular}{lll}
		\hline
		Seeding      & Type                     & Hollow glass spheres       \\ 
		&                          &                                                 \\
		Light sheet  & Laser type               & Nd:YLF                                          \\
		& Wave length              & 527 nm                                          \\
		& Frequency                & 1400 Hz                                         \\
		&                          &                                                 \\
		Camera       & Model                     & Phantom v641 ($\times 3$)                               \\
		& Resolution         & $1800 \times 2560$ px                                  \\
		& Pixel size               & 10 $\mu$m                                    \\
		& Lens                     & \begin{tabular}[c]{@{}l@{}}Nikor 50 mm ($\times 3$; $2\times$ planar PIV + PLIF)\\ Nikor 105 mm ($2 \times$ planar PIV) + Nikor 60 mm (PLIF)\end{tabular} \\
		&                          &                                                 \\
		Imaging      & Viewing area      &       \begin{tabular}[c]{@{}l@{}} $300.6 \times 303.9$ mm \\ $152.6 \times 137.3$ mm\end{tabular}     \\
		& $f_{aq}$                 & 50 Hz                                                \\
		& $ \dt t $ (PIV)             &  1250, 800 $\mu$s                              \\
		& $N_{aq}$                 & 2537                                            \\
		& Spatial resolution (PIV) &  2.85, 1.29 mm                                \\
		&                          &                                                 \\
		PIV analysis & Interrogation area       & $24 \times 24$ px                               \\
		& Window overlap                  & 50\%                                            \\
	\end{tabular}
\end{table}

Commonly, the instantaneous extent of a turbulent flow (wake) in a non-turbulent background can be defined by identifying the bounding surface (irrotational boundary) of the  turbulent/non-turbulent interface by using a threshold based on enstrophy, $\omega^2/2$ where $\omega_i$ is the vorticity vector, \citep{DaSilva2014} since the irrotational boundary is defined as an isosurface of zero enstrophy \citep{Corrsin1955}.
Since the aim of the present study is to observe the near-wake behaviour in the presence of background turbulence, it is not possible to rely on enstrophy as a marker to identify the parts of fluid that constitute the cylinder wake since $\omega^2 \neq 0$ on both sides of the turbulent/turbulent interface.
Following the methodology of \citet{Kankanwadi2020,Kankanwadi2022} we seed the wake with a passive scalar of high Schmidt number, $ Sc $, which represents the ratio of the momentum diffusivity to the mass diffusivity. 
A high $ Sc $ ensures that diffusion is restricted to exceptionally small length scales meaning that the scalar faithfully tracks the fluid in the wake. 
Accordingly, Rhodamine 6G was injected iso-kinetically into the wake from the rear face of the cylinder via a micro dosing pump (B\"{u}rkert micro dosing unit 7615) operated at 10 Hz. 
This pump discretely released 5 $\mu$l of fluid per stroke and the flow output was smoothed by routing the released fluid through a 2 m long elastic tube that connected the pump to the release hole in the back of the cylinder. 
Rhodamine 6G has a Schmidt number of approximately 2,500 in water \citep{Vanderwel2014}, making it an ideal candidate for use in this experiment. 
Tests conducted in a non-turbulent background examining the very near field of the wake that compared the extent of the scalar to the enstrophy distribution, confirmed that the released scalar was well stirred and faithfully marked the entire extent of the wake fluid \citep{Kankanwadi2020}.

Two separate combined planar PIV + PLIF experiments were conducted in which two PIV cameras (Phantom v641s) were used in order to image an extended field of view in combination with one PLIF camera (also a Phantom v641).
These experiments corresponded to the case in which the circular cylinder did and did not have a splitter plate attached to its rear surface.
Table \ref{tb: exp parameters} gives the details of these fields of view as well as all the other pertinent experimental parameters.
The PLIF experiment exploits the shifted emission peak of Rhodamine 6G to allow the PLIF camera, along with the help of a low pass filter, to capture the tracer data whilst ignoring any light scattered by seeded PIV particles. 
Rhodamine 6G has an absorption peak of 525 nm with an emission peak at a wavelength of 560 nm \citep{Arcoumanis1990}. 
The PIV camera was thus fitted with a bandpass filter centred around 532 nm, to prevent any reflected light from the dye affecting the PIV measurements. 
Both fields of view imaged the near-wake region, encompassing the flow immediately downstream of the cylinder.

%Figure \ref{fig: exp setup} and table \ref{tb: exp parameters} depict the experimental setup and provide a summary of the experimental parameters respectively. 
Illumination for both the PIV and PLIF experiment was provided by a Nd:YLF high-speed laser operating at 1400 Hz. 
Hollow glass spheres were mixed into the flow to be used as seeding particles for the PIV experiment. 
The nominal diameter of these particles is approximately 10 $\mu$m. 
Calculating the particle response time based on Stokes flow ($ \tau_p $) equates to 6 $\mu$s. 
For the particles to faithfully follow all the scales of fluid motion then the time scale with respect to the smallest eddies in the flow, the Kolmogorov time scale $ \tau_\eta $, needs to be such that the Stokes number $ St_p = \tau_p/\tau_\eta \ll 1 $. 
The resulting particle Stokes number for the worst case scenario (a group 3 case) was $St_p= 1.9\times10^{-4} $, but for all cases $\mathcal O(10^{-5}) \lesssim St_p \lesssim \mathcal O(10^{-4})$ indicating that the chosen seeding particles were appropriate.

Post-processing of all collected PIV images was done through a multi-pass cross-correlation method, which is implemented in the commercial PIV software offered by LaVision, DaVis. 
A total of three passes with a reducing window size were used to produce instantaneous velocity vectors. 
Images were also oversampled so that the overlap between two windows was equal to 50\%. 
A minimum peak ratio between the primary and secondary correlation peaks of 1.2, along with a median filter, were applied to reject spurious vectors. 
Any missing vectors were then interpolated. 
For all runs, the experimental PIV settings were optimised in order to keep the amount of replaced vectors below 3\%. 
Post-processing for the PLIF experiment was done using in-house MATLAB codes.

\begin{table}
	\centering
	\caption{Turbulence-generating-grid - cylinder spacings used to produce the experimental envelope illustrated in figure \ref{fig: Near wake exp envelope}. $L_0$ is the mesh size for the largest cell in the grid, SFG refers to a space-filling fractal square grid whilst SRG refers to a regular grid. For further geometric details about the turbulence-generating grids used see \citet{Kankanwadi2020}.}
		\label{tb: exp parameters}
		\begin{tabular}{lll|lll}
			\hline
			 & & & \multicolumn{3}{|c}{Distance $x_0$ to centre of FoV} \\ \cline{4-6} 
			Case & Grid & $L_0$ & $x_0$ (mm) & $x_0/L_0$ & $x_0/d$     \\ \cline{1-6}
			Run1a &	No Grid	&           &     &		&	\\
			Run2a &	SFG7\_1060 &	353.5 &	1060 &	3.0	& 35.3 \\
			Run2b &	SRG38\_493 &	38 &	493 &	13.0 &	16.4 \\
			Run2c &	SFG5\_2496 &	311.9 &	2496 &	8.0	& 83.2 \\
			Run2d &	SFG5\_3535 &	311.9 &	3535 &	11.3 &	117.8 \\
			Run2e &	SFG7\_2121 &	353.5 &	2121 &	6.0	& 70.7 \\
			Run3a &	SRG111\_750 &	111 &	750 &	6.8	& 25.0 \\
			Run3b &	SRG111\_602 &	111 &	602 &	5.4	& 20.1
		\end{tabular}
	\end{table}

\subsection{Wake identification methodology}

\begin{figure*}
	%\hspace{-0.5cm}
	\begin{tikzpicture}
	\node[anchor= west,inner sep=0] (0) at (0,1.27)
	{\includegraphics[width = 0.3\textwidth]{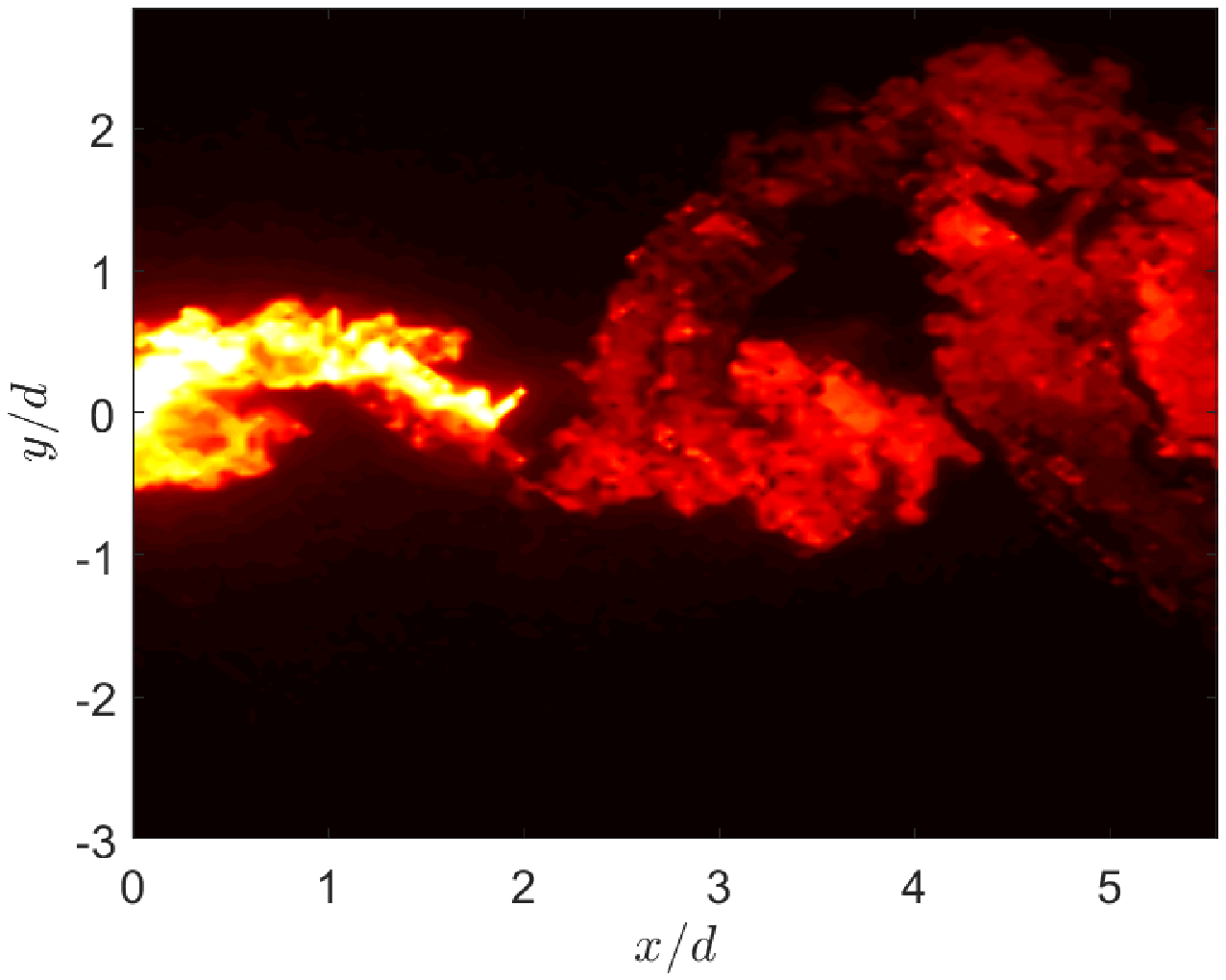}
		\includegraphics[width=0.3\textwidth]{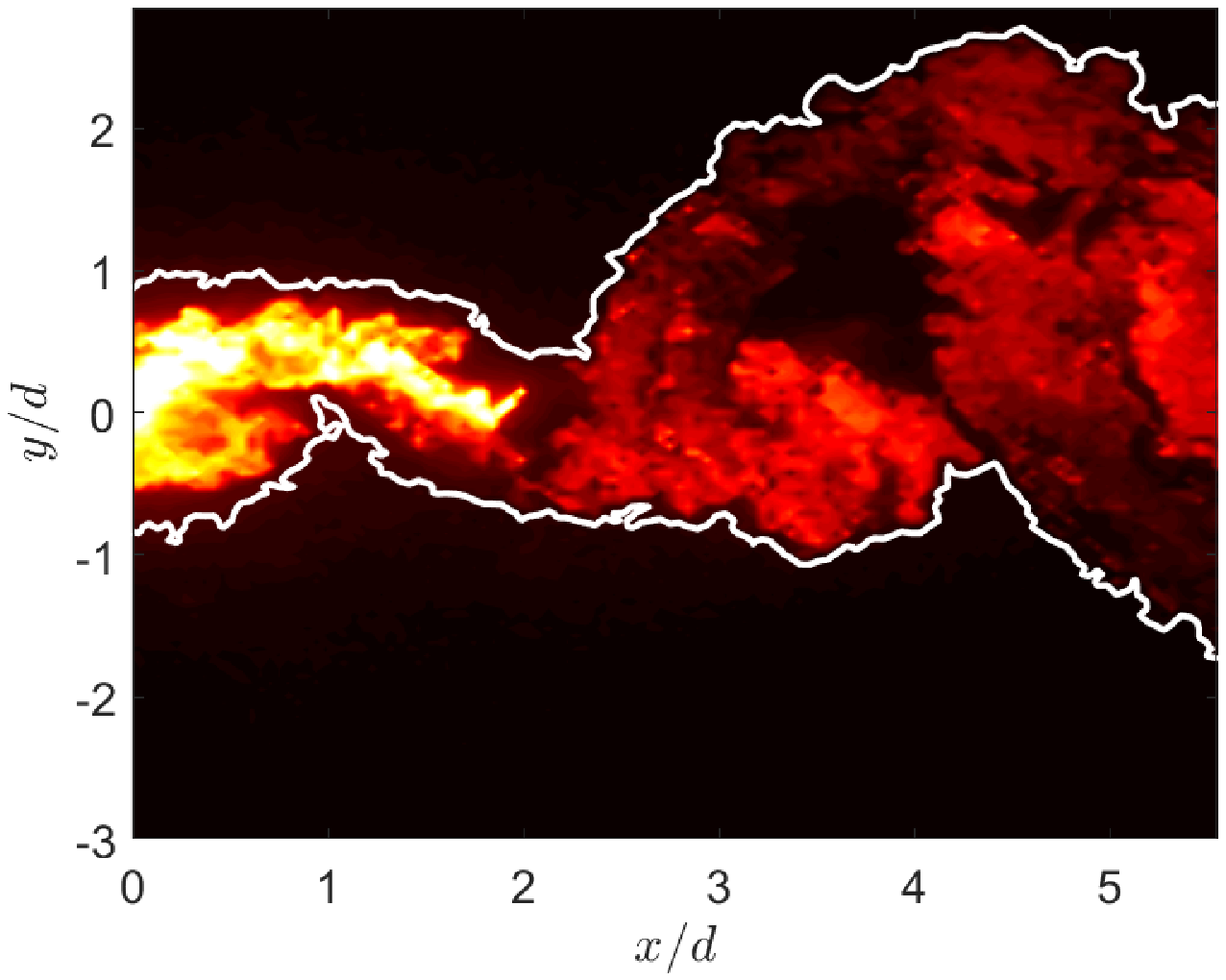}
		\includegraphics[width=0.3\textwidth]{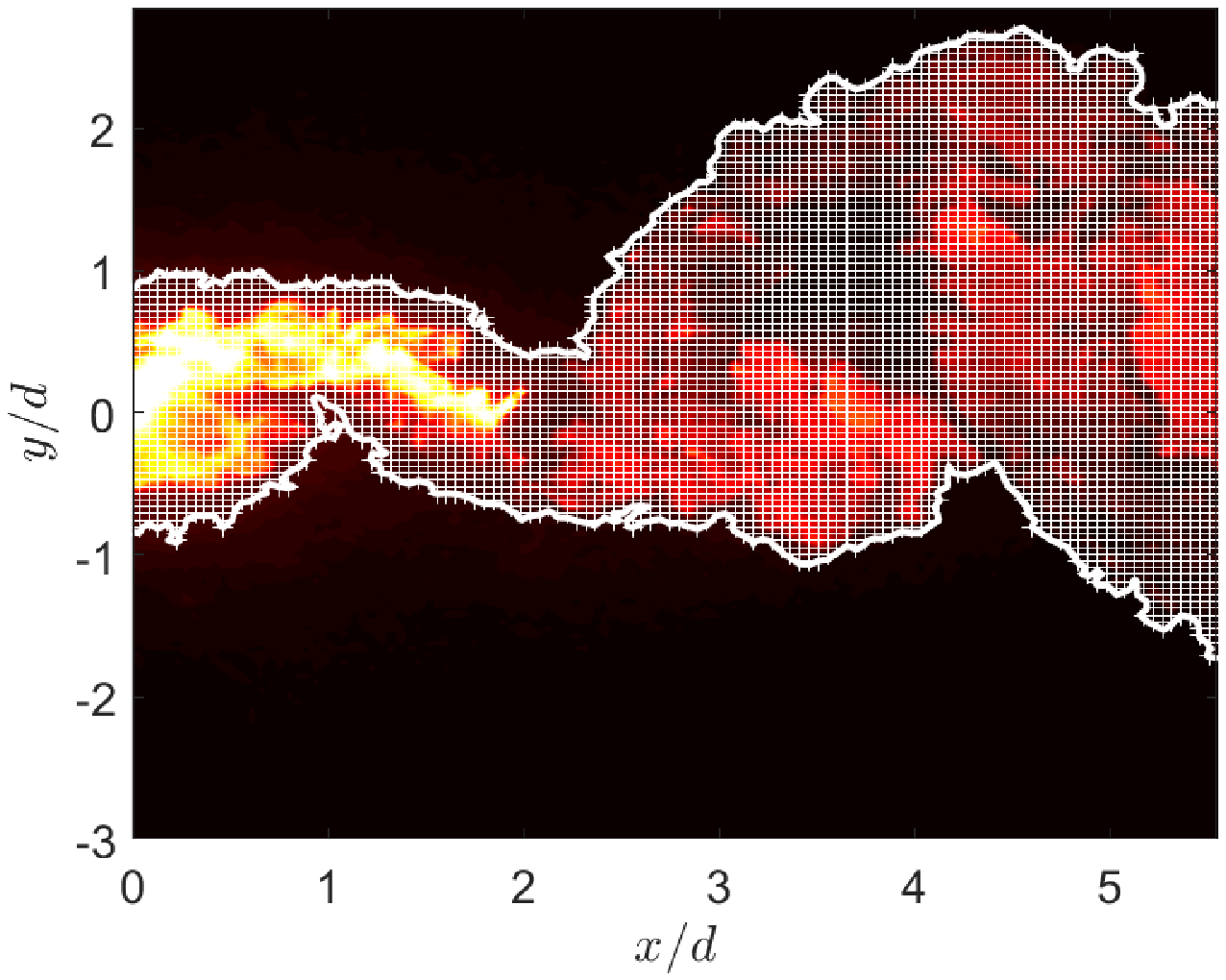}};
	\node (0) at (0,2.6) {(\textit{a})};
	\node (1) at (5.8,2.6) {(\textit{b})};
	\node (2) at (5.8*2,2.6) {(\textit{c})};
	\end{tikzpicture}
	\caption{(\textit{a}) A typical PLIF image. (\textit{b}) The solid white lines represent the wake boundary as identified by the light-intensity gradient metric. (\textit{c}) All points within the identified contours are considered to be ``in-wake''.}
	\label{fig: Wake ident method}
\end{figure*}

The methodology used to identify fluidic regions that belonged to the near wake is based on a similar light-intensity gradient based scheme used previously in \citet{Kankanwadi2020,Kankanwadi2022}. 
A threshold on $| \nabla \Phi |$, where $\Phi$ is the pixel-by-pixel light intensity captured by the PLIF camera, is determined based on monitoring the mean enstrophy in the data points being identified as out-of-the-wake.
The chosen value of the threshold is where there is a plateau in this metric, suggesting that small changes in the threshold do not alter the position of the turbulent/turbulent interface.
Validation of this methodology is presented in \citet{Kankanwadi22thesis}.
The gradient of $\Phi$ is chosen to avoid issues associated with the ``halo'' effect induced by secondary fluorescence surrounding concentrated regions of scalar \citep{Baj2016}, which is a particular issue in the near wake.
Further details on this methodology can be found in \citet{Kankanwadi2020}.
The current method differed by instead of identifying one continuous contour, the two longest contours in the field of view were identified and used as the boundaries of the wake.
The region that lies in between the two contours is classified to be inside the wake (later referred to as ``in-wake''). 
Figure \ref{fig: Wake ident method} depicts this process pictorially. 
This process was repeated for all snapshots in the run and thereby, a complete map of the development of the wake in the near-field was obtained.

It is interesting to note that a large pocket of fluid that does not have a significant scalar presence is considered to be part of the wake in figure \ref{fig: Wake ident method}. This follows from the adopted methodology to consider all regions of fluid within the identified contours to be ``in-wake''. This pocket represents a region of engulfed background fluid and is a typical example of the engulfment process, which is of great importance in the near field \citep{Dahm1987,Dahm1990,Burridge2017}.
In most far-field analysis, it is possible to ignore entrainment that happens through engulfment, since the proportion of growth through engulfment may be considered negligible. \cite{Mistry2016} show that in their analysis of a jet, at a streamwise distance of 50 diameters, engulfed fluid only amounts to 0.44\% of the jet area.
Similar liberties may not be taken in the near-field since engulfment potentially has a more pronounced role in the entrainment process.

\subsection{Triple decomposition}
\label{sec: Triple decomposition}
The flow field being examined is dominated by coherent structures that are periodically shed by the cylinder. 
It is therefore useful to decompose the velocity field into a periodic and stochastic component. 
This follows the triple decomposition method outlined by \cite{Reynolds1972}. 
A part of the derivation has been recreated below for reference. 
The method is based on decomposing the fluctuating velocity component $ s'(\textbf{x},\textbf{t}) $, into, a coherent fluctuation, $ \tilde{s} $, and a stochastic fluctuation $ s'' $. 
The total variable, $ s $ is described by the sum in \eqref{eq: triple_decompostion} whilst \eqref{eq: numerical_triple_decomposition} represents its numerical implementation.
$ l $ is the chosen phase ($ \phi $) sub-population, whereas $ p $ depicts the selected time snapshot.\\

\noindent\begin{minipage}{.5\linewidth}
	
	\begin{equation}
	\label{eq: triple_decompostion}
	s =  \bar{s} + \tilde{s} + s''
	\end{equation}
	
\end{minipage}%
\begin{minipage}{.5\linewidth}
	
	\begin{equation}
	\label{eq: numerical_triple_decomposition}
	s_{\phi = l,\ n = p} =  \bar{s} + \tilde{s}_{\phi = l} + s''_{n = p}
	\end{equation}
	
\end{minipage}\\

For a population of size $ N $, with samples, $ s_n $, an average over the population can be defined in two ways, both of which are described by \eqref{eq: population_mean} and \eqref{eq: phase_average}. 
Note that $ \langle s \rangle $ represents the phase average and is evaluated as a mean over a sub-population of samples that correspond to the same phase angle. 

\begin{align}
N\bar{s} &=  \sum_{n = 1}^{N} s_n \label{eq: population_mean} \\ 
N_i\langle s \rangle &=  \sum_{\nu = 1}^{N_i} s_{\nu} \quad (i = 1,2,...,\ \label{eq: phase_average} \operatorname{Number\ of\ phase\ sub-populations})
\end{align}

\noindent The periodic component, $ \tilde{s} $ can then be calculated as,
\begin{equation}
\label{eq: periodic_component}
\tilde{s} =  \langle s \rangle - \bar{s}
\end{equation}

Complexity arises in assigning a phase value for each time snapshot of the signal. Investigators in the past have tackled this using a collection of methods.
\cite{Hussain1970} knew \emph{a priori} the instantaneous phase value of their signal, hence allowing them to sample at specific phase values. 
This method has been coined as conditional point averaging. 
Due to the requirement of having prior knowledge of the phase signal, this method has very specific use cases. 
\cite{Cantwell1983} proposed a more general method of phase averaging, often referred to as bin averaging. They made use of a pressure sensor mounted on the front face of their circular cylinder in order to record vortex shedding phase. 
It is then trivial to bin each sample into a collection of phase sub-populations. 
However, due to the lack of a pressure sensor in our experiment to provide a phase reference, a phase averaging method without this requirement is needed. 
For the purposes of this study, a method developed by \cite{Baj2015} will be utilised. 
They made use of optimal mode decomposition (OMD),  a technique developed by \cite{Wynn2013}, to evaluate modes that are associated to vortex shedding. 
An overview of the application of this method to extract the phase signal is given in appendix \ref{sec:omd}.
For a more thorough description, the reader is advised to refer to references \cite{Wynn2013,Baj2015}.

\begin{figure}
	\hspace{-0.5cm}
	\begin{tikzpicture}
	\node[anchor= west,inner sep=0] (0) at (0,1.27)
	{\includegraphics[width = 0.5\textwidth]{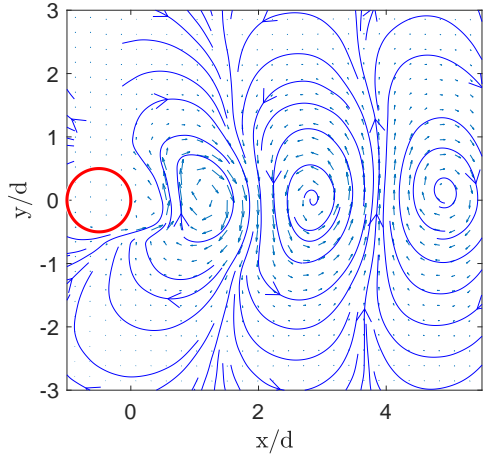}
		\includegraphics[width=0.5\textwidth]{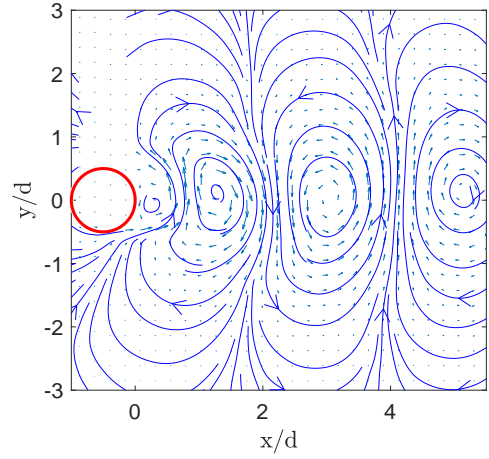}};
		\node[anchor= west,inner sep=0] (1) at (0,-6)
		{\includegraphics[width=0.5\textwidth]{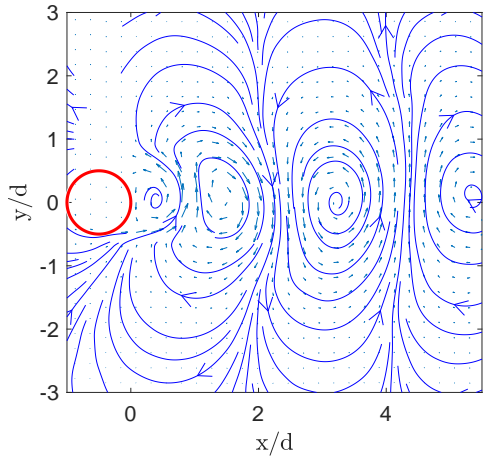}
    		\includegraphics[width=0.5\textwidth]{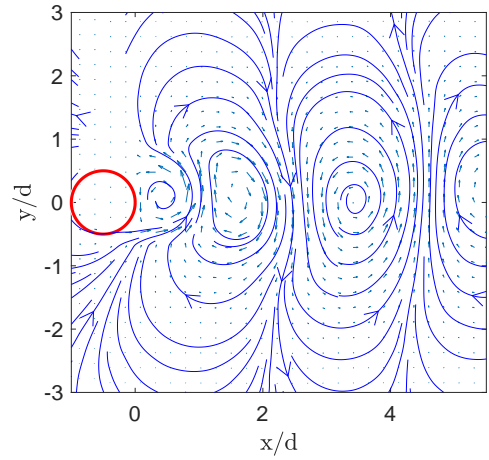}};
	\node (a) at (4.2,-2.35) {(\textit{a}) $\phi = 0.45\pi$};
	\node (b) at (12.7,-2.35) {(\textit{b}) $\phi = 0.55\pi$};
	\node (c) at (4.2,-9.7) {(\textit{c}) $\phi = 0.65\pi$};
	\node (d) at (12.7,-9.7) {(\textit{d}) $\phi = 0.75\pi$};
	%\node (a) at (0.5,3.5) {(\textit{a})};
	%\node (b) at (9,3.5) {(\textit{b})};
	%\node (c) at (0.5,3.5-6.27) {(\textit{c})};
	%\node (d) at (9,3.5-6.27) {(\textit{d})};
	\end{tikzpicture}
	\caption{Plots of phase-averaged velocity across four consecutive phase bins produced from a data set of the no-grid case. Superimposed on top of quiver plots of $\tilde{u} \, \& \, \tilde{v}$ are plots that depict streamlines.}
	\label{fig: Phase averaging}
\end{figure}

Upon establishing the instantaneous phase signal for vortex shedding, collected samples were binned into a total of 20 sub-populations. Each phase bin had a size, $ \Delta \phi = \pi/10 $. In a typical run that had a total of 2537 snapshots, each phase bin consisted of approximately 127 snapshots. Figure, \ref{fig: Phase averaging} provides typical examples of the periodic component of velocity over a few consecutive phase bins for the no-grid case. These plots depict both the formation of a vortex in the immediate vicinity of the cylinder and also the advection of shed vortices downstream of the cylinder. Comparing these results with literature \citep{Cantwell1983} confirm a successful implementation of the methodology.

\section{Wake growth}

\begin{figure}
	\hspace{-0.5cm}
	\centering
	\begin{tikzpicture}
	\node[anchor = west, inner sep=0] (0) at (3,1.27)
	{\includegraphics[width = 0.6\textwidth]{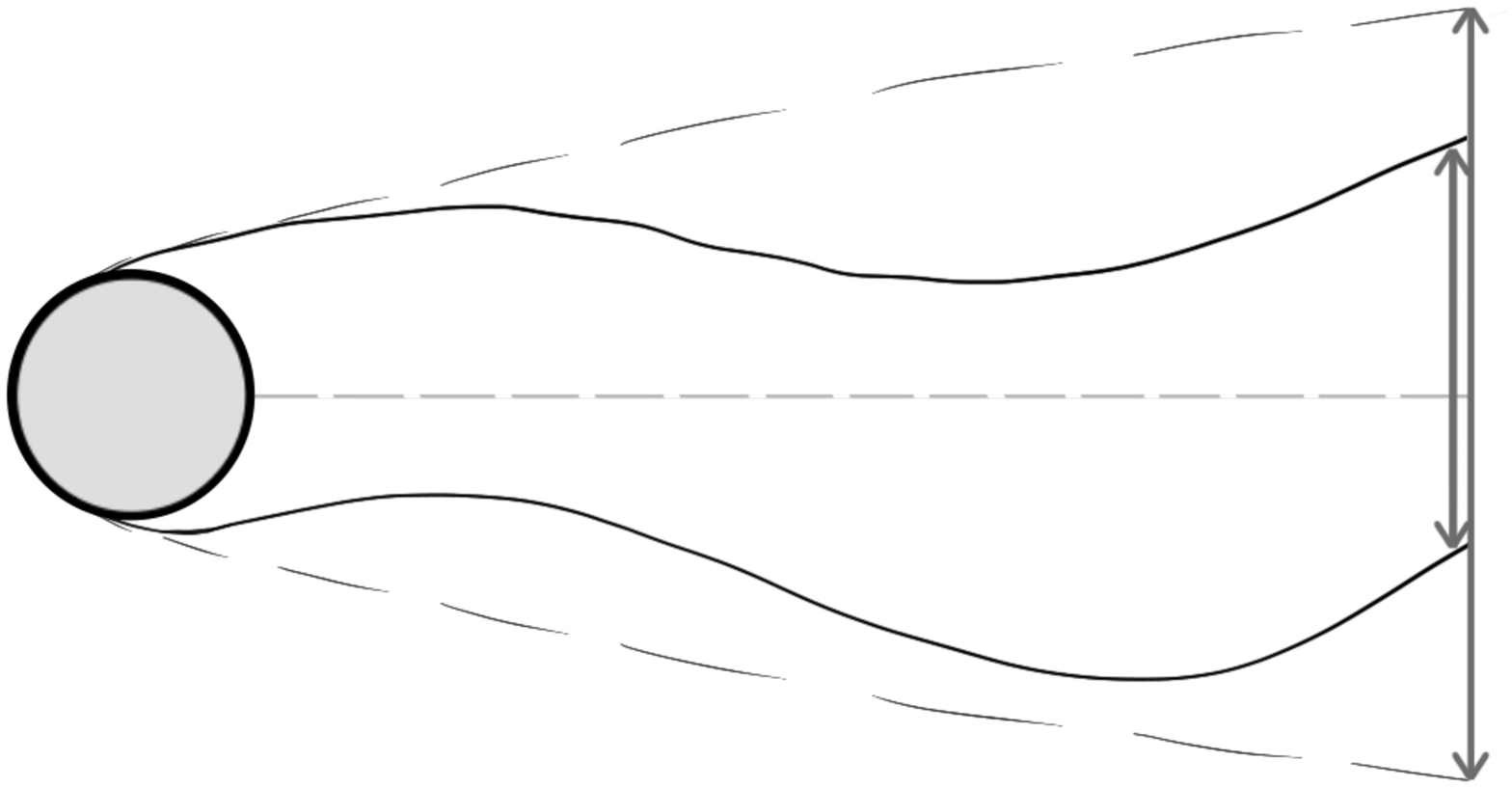}};
	\node (1) at (12.1, 1.9) {$\delta(x,t)$};
	\node (2) at (13.15, -0.5) {$\delta_\Phi$};
	\end{tikzpicture}
	\caption{A meandering wake. Cartoon highlights both the instantaneous wake width, $\delta(x,t)$, and the time averaged scalar extent of the wake, $\delta_\Phi(x)$}
	\label{fig: meandering cartoon}
\end{figure}

In a time-averaged sense, the spatial development of the wake in the near field can be separated into two distinct components. 
These can be identified as (\textit{a}) the growth of the wake itself through entrainment and (\textit{b}) the envelope of the spatial extent of the wake induced by its ``meandering'' downstream of the cylinder.
We term meandering as the transverse oscillations of the shed vortical structures.
Accordingly, relying on the mean coverage of the scalar to comment on wake growth does not present the whole picture, since the growth of a wake in the near field is an amalgamation of the true growth through entrainment and an illusion of increased wake width in a time-averaged sense due to meandering.
Figure \ref{fig: meandering cartoon} provides an illustration of a meandering wake, where both growth mechanisms are realised.
%whilst also highlighting both the traditional wake width calculated using the time averaged scalar extent as well as the instna
The following analysis aims to decouple the two phenomena to gain a clearer understanding of wake growth mechanisms when subjected to turbulent free-streams.

\begin{figure}
	\centering
	\includegraphics[width = 0.7\textwidth]{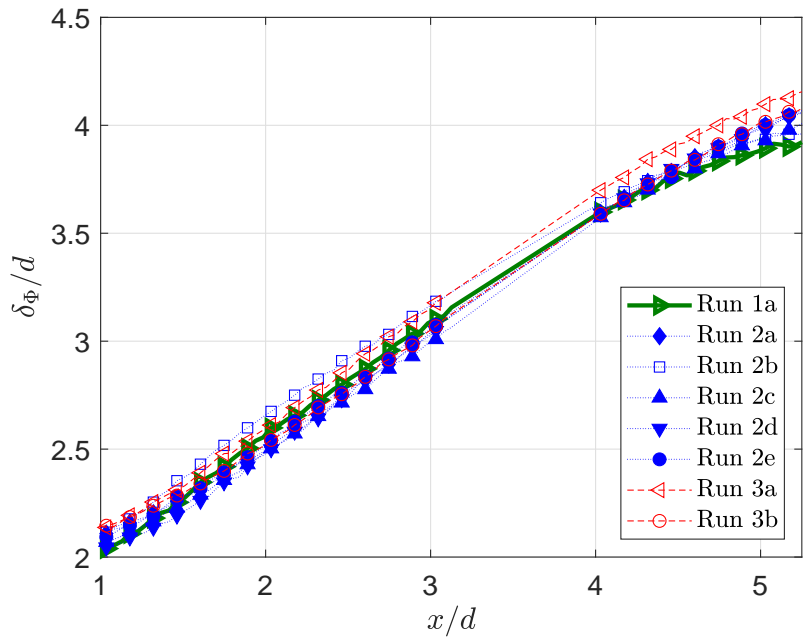}
	\caption{Time-averaged wake width obtained using a threshold on the total extent of the wake. Note that the region between $3 \lesssim x \lesssim 4$ has been omitted due to an experimental artefact (see discussion in the text).}
	\label{fig: Time Avg PLIFInWake}
\end{figure}

\begin{figure}
	\centering
	\includegraphics[width = 0.7\textwidth]{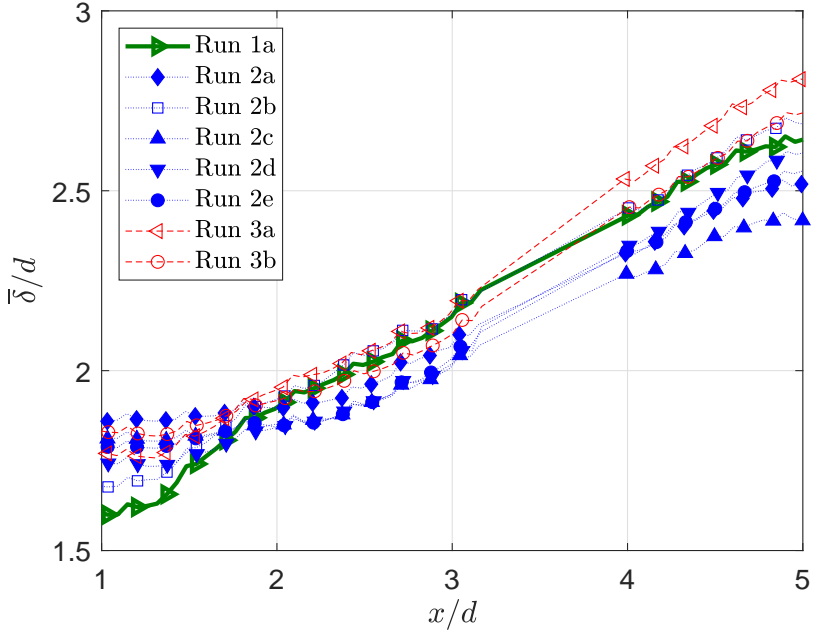}
	\caption{Temporally averaged instantaneous wake width as a function of streamwise distance downstream of the cylinder.} 
	\label{fig: Dynamic wake width}
\end{figure}

\begin{figure}
	\centering
	\includegraphics[width = 0.7\textwidth]{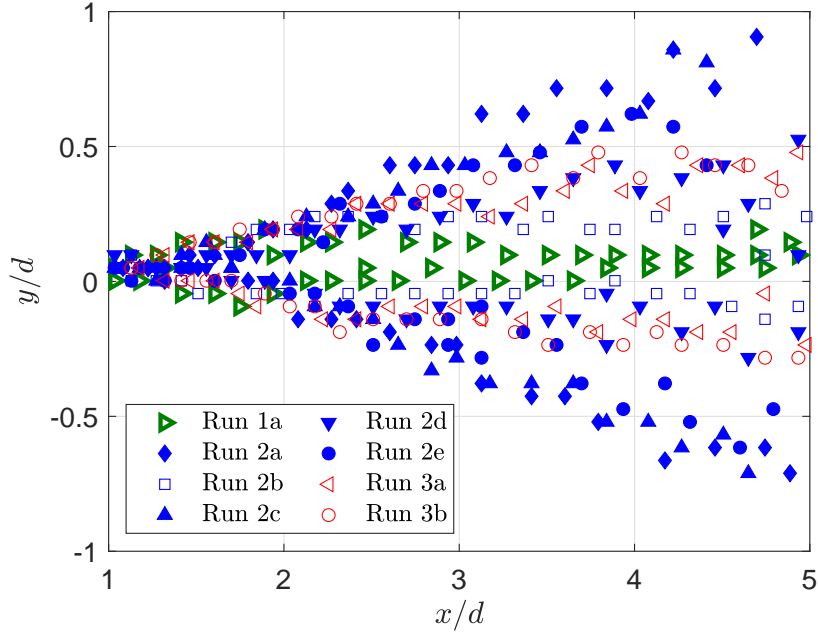}
	\caption{Vortex centres extracted from the phase-averaged velocity ($\tilde{u} \, \& \, \tilde{v} $).}
	\label{fig: Vortex centres}
\end{figure}

However, prior to this distinction being made, the wake growth is presented as a time average of the extent of the wake. 
This is important to include, since most literature refers to wake growth in this manner, e.g. empirical wind-turbine wake growth models \citep{Jensen1983,Bastankhah2014}.
In order to evaluate this metric, the time average of the spatial extent of the wake is first calculated. 
A user defined threshold is then applied to the time-averaged field in order to establish the wake width.
The chosen threshold represented a probability of 18\% that a given point was considered to be within the wake over the time series.
The chosen threshold probability of 18\% is arbitrary, nevertheless it was ensured that the results were independent of the chosen threshold following a sensitivity study, see Appendix D1 of \citet{Kankanwadi22thesis}.
A methodology such as this is acceptable since the aim is to assess the trends of various runs with respect to each other as opposed to identifying a quantitative value for the time-averaged wake width (in this sense).
Figure \ref{fig: Time Avg PLIFInWake} depicts the result of this process, illustrating the time-averaged wake width established using our defined ``in-wake'' threshold, $\delta_{\Phi}$ from figure \ref{fig: meandering cartoon}, as a function of streamwise distance $x$.
The figure establishes that an increase in this metric is observed for cases where the wake is subjected to freestream turbulence. 
At $x/d = 5$, group 3 cases display the largest wakes, followed by cases in group 2 and the baseline no-grid case presents a wake with the least width. This result is consistent with the traditional understanding of increased entrainment and mixing in the near-field due to background turbulence \citep{Bearman1983}, hence resulting in wider wakes, something that has been adopted in analytical wind-turbine wake modelling strategies \citep[e.g.][]{Niayifar2016}.
We discuss the behaviour of the various cases for $x/d < 5$ later on in this section, in conjunction with the results subsequently illustrated in figures \ref{fig: Dynamic wake width} and \ref{fig: Vortex centres}.

Before progressing to the decoupled analysis, the lack of data between streamwise positions of $3 \lesssim x/d \lesssim 4$ must be addressed. 
Data in this region has been deliberately omitted from the analysis to prevent an experimental artefact from unduly influencing the presented results. 
The artefact arises from the presence of bright reflections in the background of images taken by the PLIF camera. 
Analysis that relies on the PLIF experiment are therefore compromised and hence data collected in this region is omitted from any further analysis. 
Attempts to remove the background reflection using image processing, whilst promising, were ultimately unsuccessful.

Wake growth through entrainment (in the time-averaged sense) is investigated by examining the temporal mean of the instantaneous wake width, $\overline \delta (x)$. 
This is the time average of the metric $ \delta(x,t)$ from figure \ref{fig: meandering cartoon} and is depicted in figure \ref{fig: Dynamic wake width} as a function of streamwise distance downstream of the cylinder. 
This result is intriguing since background turbulence is seen to both promote (group 3 cases), and also suppress wake width (group 2 cases with the exception of run 2b). 
Close inspection reveals that there is a link to the length scale of the subjected background turbulence as all of the runs with suppressed wake width in the near field have a large integral length scale in the background turbulence (runs 2a,c,d,e) whilst run 2b (with a smaller background $L_{12}$) tracks the no-grid case as well as run 3b.
The slope $\dt \overline \delta / \dt x$ of the time-average of the instantaneous wake width is indicative of the entrainment rate into the wake.
Should wake growth be simplified to a linear model, $\overline \delta / d = m (x/d) + c$, $m$ may be considered to represent the entrainment rate whilst $c$ accounts for other near-wake effects, in effect imposing a near-wake ``boundary condition'' on the linear wake-spreading region.
The gradient is estimated by applying a linear regression for points that lie between $2.5<x/d<5$. 
For these $x$-positions, group 3 cases present the largest mean gradient ($m = 0.32$, averaged over run 3a and run 3b), whereas the gradient for the no-grid case ($m = 0.27$) is equivalent to group 2 ($m = 0.26$ averaged over all group 2 cases with run 2e, that with the highest TI, having the largest individual value of $m$).

Our observation that the presence of background turbulence enhances the entrainment rate, and the influence of integral length scale of the freestream turbulence on $\overline \delta (x)$, are in contrast to the results from \citet{Kankanwadi2020} from the far-wake region ($x \approx 40d$) of a turbulent cylinder wake.
Here it was observed that there is an inverse correlation between turbulence intensity of the freestream turbulence and the entrainment mass flux, i.e. high freestream turbulence intensity (e.g. group 3 cases) led to a suppression of entrainment rate.
This was driven by an increased prevalence of intermittent but powerful detrainment events.
Additionally, the integral length scale of the freestream turbulence was shown to have no correlation to the entrainment rate into the wake.
In the far wake the large-scale coherent structures have died out, so we postulate that integral length scale of the freestream turbulence has a stronger effect on the near wake due to the presence of large-scale (von K\'{a}rm\'{a}n) vortices.
\citet{Yule1978} identified that large-scale engulfment is the dominant entrainment mechanism in the near field of a turbulent jet, whereas small-sale nibbling is the dominant entrainment mechanism in the far field, accounting for 90-93\% of the entrained mass flux \citep[e.g.][]{Westerweel2005}.
We therefore further postulate that the presence of background turbulence has the effect of promoting engulfment whilst it suppresses nibbling (via promoting intermittent detrainment events).
Accordingly, in the near field of turbulent shear flows where engulfment is the dominant entrainment mechanism this leads to an increased entrainment rate (relative to a non-turbulent background) whilst in the far field this leads to a reduction in the entrainment rate.
We therefore finally postulate that there exists a ``crossover'' point at which the presence of freestream turbulence switches from enhancing entrainment to suppressing it in relation to a non-turbulent background.

\begin{figure}
	\hspace{-0.5cm}
	\begin{tikzpicture}
	\node[anchor= west,inner sep=0] (0) at (0,1.27)
	{\includegraphics[width = 0.5\textwidth]{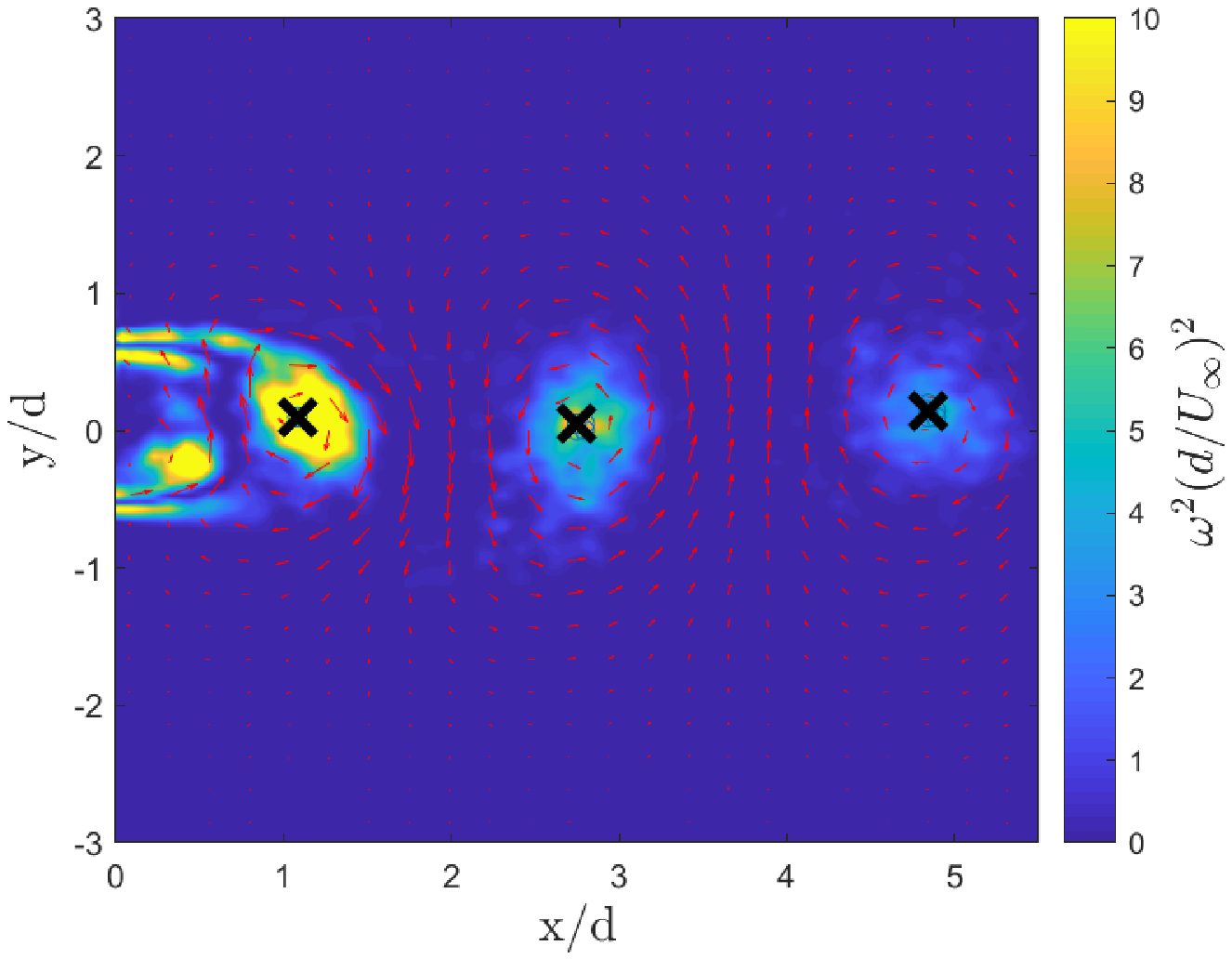}
		\includegraphics[width=0.5\textwidth]{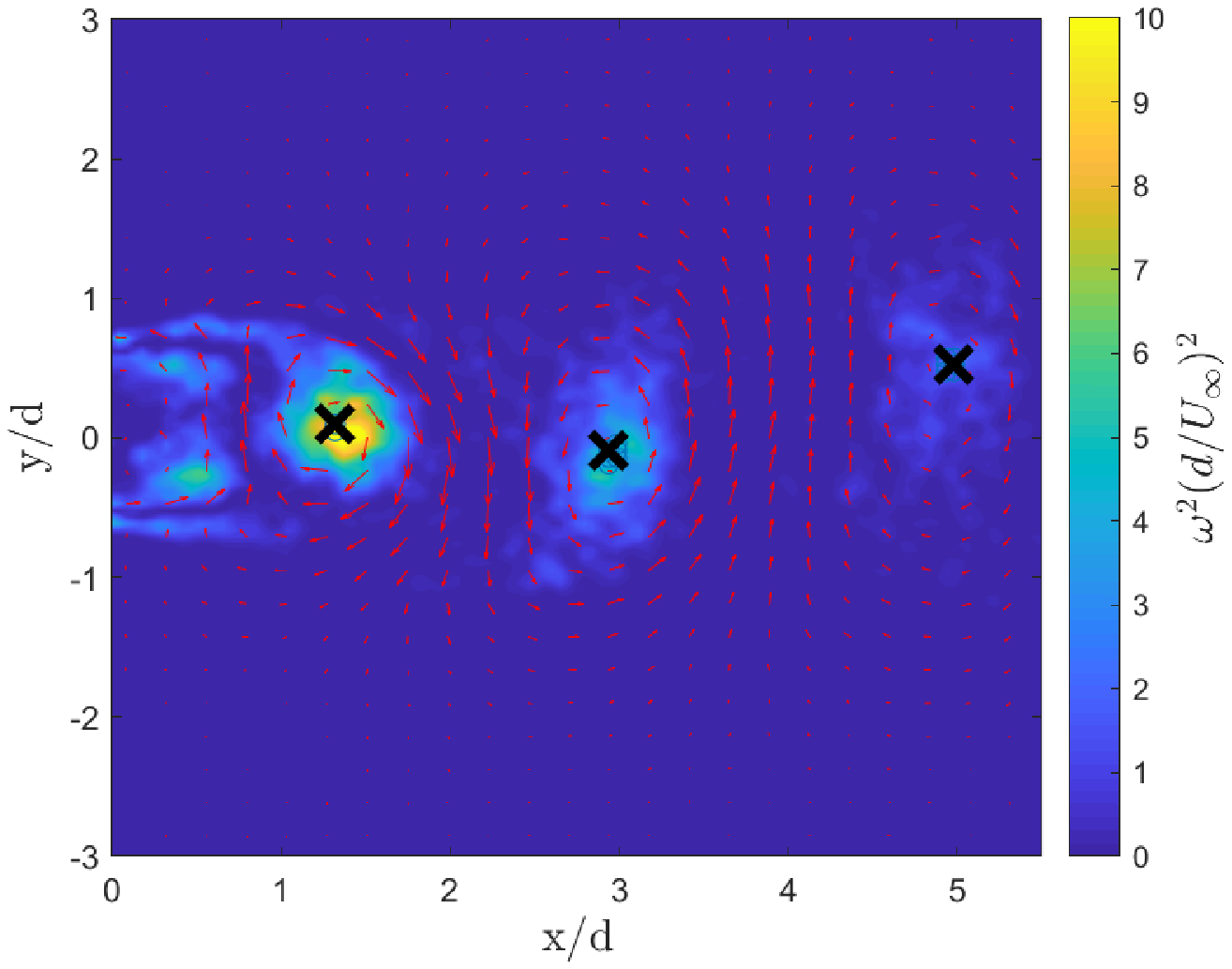}};
	\node (a) at (4.2,-2.35) {(\textit{a}) Run 1a - No-grid case};
	\node (b) at (12.7,-2.35) {(\textit{b}) Run 3a};
	\end{tikzpicture}
	\caption{Plots depict typical phase-averaged velocity fields superimposed on top of the underlying enstrophy field. Identified vortex centres are highlighted with the black cross.}
	\label{fig: vortex tracking}
\end{figure}

Background turbulence is also seen to have a defining effect on the ``meandering'' observed in the near-wake. 
Figure \ref{fig: Vortex centres} aims to quantify this effect by plotting the paths taken by vortex centres as they are advected downstream of the cylinder. 
To achieve this, vortices were tracked and analysed in a phase-averaged sense. 
For each phase bin, vortex centres were identified and tracked by applying a threshold on enstrophy and velocity magnitude on the periodic component of fluctuating velocity, $\tilde{u}$ \& $\tilde{v}$. 
%Such an approach was possible since the phase averaging significantly reduced the level of noise. 
Thresholds ensured that regions of highest enstrophy combined with the lowest velocity magnitude were appropriately identified. 
A clustering scheme was then employed over the identified points, allowing for the centre of the vortex to be evaluated.
Figure \ref{fig: vortex tracking} illustrates the methodology. 
Depicted are a couple of typical phase-averaged velocity fields in the form of red vectors for two different cases. 
It should be noted that not all vectors are plotted in the interest of clarity. 
The quiver plot is superimposed on top of a contour plot of the enstrophy field. 
Following the application of the vortex identification methodology described above, the identified vortex centres are highlighted in the figure.
These results were also compared with the approach outlined by \cite{Vollmers2001}, which involved examining the $\lambda_2$ criterion for two-dimensional data. 
\cite{Vollmers2001} explains that vortices appear for non-real eigenvalues of the velocity gradient tensor, $\mathcal{G}$. 
The discriminant, $\lambda_2$, separates vortices from other structures. 
Mathematically, this is defined as, $\lambda_2 = (\mathrm{trace} \: \mathcal{G})^2 - 4 \: \mathrm{det}(\mathcal{G})$. 
Regions where this metric is negative represent vortical structures. 
Vortex centres can then be identified as the centroid of each region where $\lambda_2 < 0$. Results using both methods were found to be similar.

Identified vortex centres are depicted as individual data points in figure \ref{fig: Vortex centres}. 
It is clear to see that freestream turbulence significantly increases the transverse location relative to the centreline of these coherent structures. 
This leads us to the conclusion that applied freestream turbulence increases the level of meandering of the wake. 
It is also interesting to note that length scale of the background turbulence yet again plays a part; all of the cases that display heavily increased meandering originate from runs that have an increased integral length scale. 
Run 2d is an outlier as it displays only a slight increase in meandering despite the background turbulence housing a large integral length scale.

The two decoupled sets of analysis may be reconciled with the traditional wake width ($\delta_\Phi$) result presented in figure \ref{fig: Time Avg PLIFInWake}.
In order to do this it is helpful to re-categorise the runs with respect to $L_{12}$. 
Runs in group 3 and run 2b may be classified as runs that have a small background integral length scale, whereas the remaining runs in group 2 can be classified as runs with large background length scale (see figure \ref{fig: Near wake exp envelope}). 
In general, it is found that cases with background turbulence of a small integral length scale do not suffer from a reduced instantaneous wake width caused by the presence of a large $L_{12}$ in the background turbulence, benefit from an enhanced entrainment rate into the wake yielded by the presence of the background turbulence, and see a slightly enhanced amplitude of wake meandering relative to the no-grid case.
Accordingly, the mean wake width as defined by $\delta_\Phi$ is larger everywhere than for the case with no grid.
Cases with a large background length scale, on the other hand, are seen to display a reduction in instantaneous wake width. 
However, this is compensated for by the significant increase in the level of meandering.
In the very near-wake region, say $x/d \lesssim 3$, the increased meandering is not sufficient to overcome the suppressed instantaneous wake width meaning that $\delta_\Phi$ for these cases is smaller than the no-grid case.
However, further downstream (say $x/d \gtrsim 4$) the enhanced entrainment rate and wake meandering become sufficiently large that $\delta_\Phi$ for these cases overtakes the no-grid case.
We point out that there is a fine balance in these competing effects: the enhanced entrainment rate in the presence of background turbulence, the suppression of instantaneous wake width and the enhanced wake meandering both driven by large-$L_{12}$ background turbulence are in a fine balance with one another: the differences between $\delta_\Phi$ for the various cases in figure \ref{fig: Time Avg PLIFInWake} are relatively smaller than those in the various constituents illustrated in figures \ref{fig: Dynamic wake width} and \ref{fig: Vortex centres}.
Similarly to estimating $\dt \overline \delta / \dt x$ from the data in figure \ref{fig: Dynamic wake width} we make a linear fit to the data in figure \ref{fig: Time Avg PLIFInWake} in the range $2.5 < x/d < 5$ such that $\delta_\Phi / d = m (x/d) + c$.
We again report the value of $m$ for the no-grid case ($m = 0.44$), the mean slope for the group 2 cases ($m = 0.49$) and the mean slope for the group 3 cases ($m = 0.50$).
This again indicates the fine balance for the evolution of the mean wake width as defined in the classical way, although with a tendency for a faster wake growth in the presence of background turbulence.

\begin{figure}
	\hspace{-0.5cm}
	\begin{tikzpicture}
		\node[anchor= west,inner sep=0] (0) at (0,1.27)
		{\includegraphics[width = 0.5\textwidth]{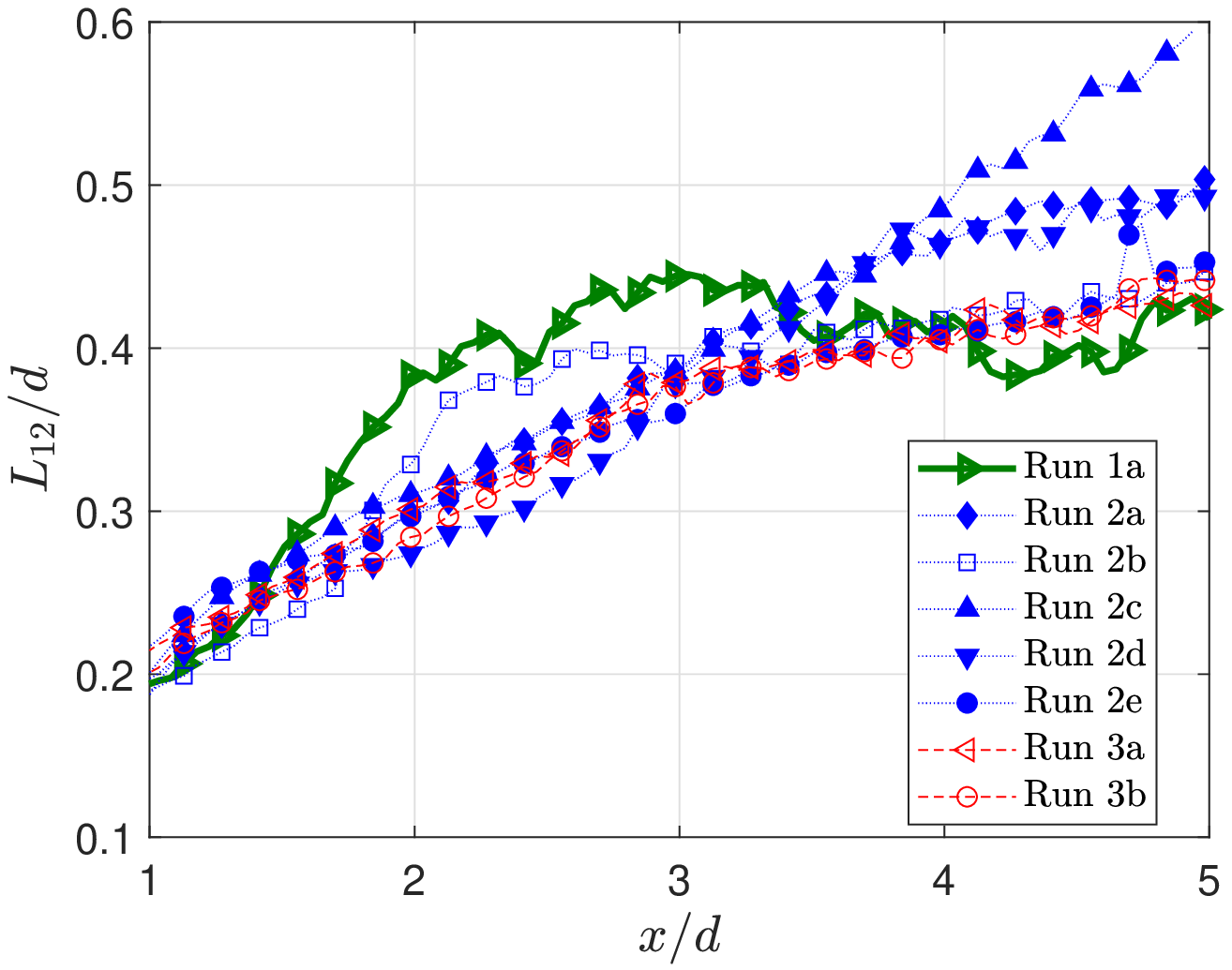}
			\includegraphics[width=0.5\textwidth]{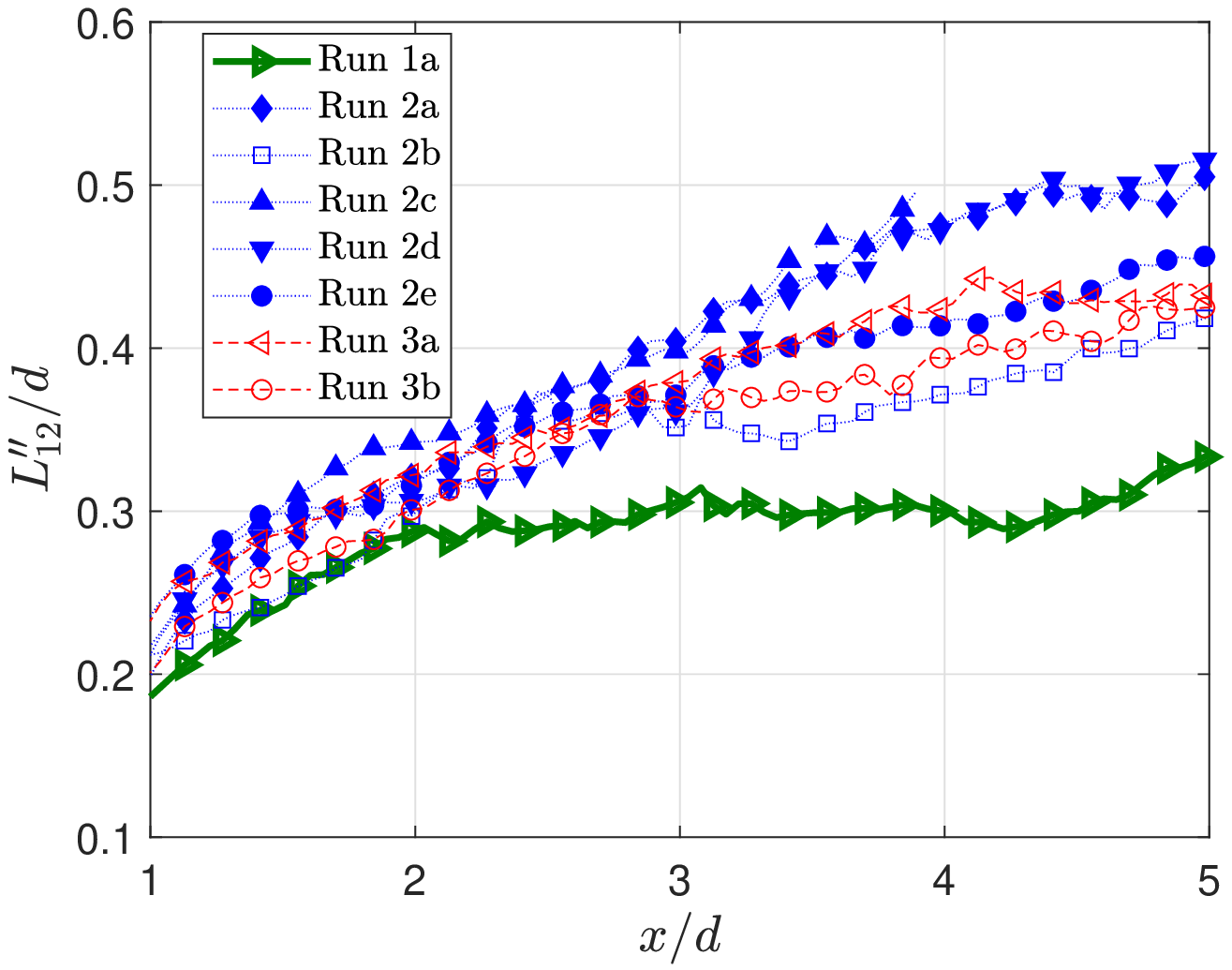}};
		\node (a) at (0,3.65) {(\textit{a})};
		\node (b) at (8.5,3.65) {(\textit{b})};
	\end{tikzpicture}
	\caption{Spatial evolution of the integral length scale within the wake computed from a fixed probe along the centreline. (a) $L_{12}$, i.e. length scale computed from the correlation of $u' = \tilde u + u''$ (b) $L_{12}''$, i.e. length scale computed from the stochastic fluctuation only $u''$.} 
	\label{fig:lscale}
\end{figure}

We conclude this section by presenting the spatial evolution of the integral length scale computed from within the wake, $L_{12}$, in figure \ref{fig:lscale}(\textit{a}).
The in-wake length scale is computed in precisely the same fashion as for $L_{12}$ for the background turbulence, with the fixed probe location being at $(x,0)$, i.e. along the centreline.
The first observation to be made is that the size of the in-wake $L_{12}$, for all cases, is small in comparison to the mean instantaneous wake width presented in figure \ref{fig: Dynamic wake width}.
This suggests that the turbulence within the wake remains distinct from that within the background since the velocity evidently de-correlates over a distance much shorter than the wake half width.
Initially, i.e $x/d \approx 1$, the in-wake integral scales for all cases, including the non-turbulent background, are similar, however the no-grid case sees the fastest initial growth rate before saturating (within the field of view explored).
All cases with a turbulent background see similar behaviour, with an initially extremely similar growth rate, before eventually saturating.
The streamwise position where this saturation occurs appears to depend on the length scale of the background turbulence, with cases from the smaller background-$L_{12}$ grouping saturating earlier, and hence at a lower value of in-wake-$L_{12}$, the only exception being run 2e.
Whilst the size of the in-wake eddies is significantly smaller than $\delta_\Phi$ there is a correlation between the larger in-wake eddies and the largest amplitude of meandering, illustrated in figure \ref{fig: Vortex centres}.
We conclude that freestream turbulence with a larger integral scale leads to the evolution of eddies with a larger characteristic size themselves within the wake.
These remain distinct eddies, though, since the velocity fluctuations de-correlated over a distance significantly shorter than the wake half width.

$L_{12}$ is computed from the total velocity fluctuation, i.e. $u' = \tilde u + u''$.
In the near-wake region explored, $1 \leq x/d \leq 5$, the velocity fluctuations contain a significant contribution from the large-scale coherent motions.
To identify whether the in-wake length scale $L_{12}$ is dominated by these coherent motions, or the stochastic turbulent fluctuations $u''$ we report the spatial evolution of the in-wake length scale $L_{12}''$, which results from the transverse spatial correlation of the in-wake $u''$ velocity fluctuations in figure \ref{fig:lscale}(\textit{b}). 
Unlike in figure \ref{fig:lscale}(\textit{a}) it can be seen that at all streamwise locations, including $x/d=1$, the cases with background turbulence have a larger stochastic turbulence length scale than the no-grid case.
For all cases with background turbulence $L_{12}''$ continues to grow throughout the domain explored and there is a tendency for cases with a larger integral length scale in the background turbulence to induce larger $L_{12}''$ within the wake.
Finally, the size of the in-wake $L_{12}''$ is similar to the in-wake $L_{12}$ for all cases with background turbulence whilst it is substantially smaller for the no-grid case.
This is symptomatic of weaker in-wake coherence for cases with background turbulence present.

\subsection{Splitter plate experiment}\label{sec: Splitter plate}

\citet{Kankanwadi2020} previously showed that the presence of freestream turbulence in the far wake led to a reduction in the mean entrainment rate in comparison to a non-turbulent background, which is the opposite conclusion to that which we now obtain in the near wake. 
The primary difference between the near wake and far wake is the decay of the large-scale coherent motions; from being the energetically dominant structures in the near wake to nearly negligible in the far wake.
Previous studies have shown that in the near field of a turbulent shear flow engulfment (driven by large-scale coherent motions) is the dominant entrainment mechanism \citep[e.g.][]{Yule1978} whilst small-sale nibbling is dominant in the far field \citep[e.g.][]{Westerweel2005}.
We previously postulated that the contrasting results between the near wake (background turbulence enhances entrainment) and the far wake (background turbulence diminishes entrainment) are due to the transition from entrainment being engulfment-driven in the near wake to nibbling-driven in the far wake.
A control experiment was therefore devised whereby we could try to artificially suppress the large-scale coherent motions in the near wake.
This control experiment thereby tries to de-couple the influence of the large-scale coherent (von K\'{a}rm\'{a}n) vortices, which drive both the wake meandering and large-scale engulfment, from ``incoherent'' small-scale entrainment, i.e. nibbling.
To this end, a splitter plate was designed and manufactured to prevent the two shear layers, shed from opposite sides of the cylinder, from immediately interacting.
Details of the design of the splitter plate, and validation that it successfully eliminated the large-scale coherence in the wake of the combined cylinder - splitter plate system are provided in appendix \ref{sec:split}.

Figure \ref{fig: Splitter wake width} outlines the temporal average of the instantaneous wake width downstream of the cylinder with an attached splitter plate.
There are two different data sets, an upstream one and a downstream one.
It should be noted that the (upstream) data for $x/d < 6.59$, the splitter plate length, is calculated differently to that from the downstream data set.
In the upstream data set, only one half of the wake is visible to the cameras, since the splitter plate casts a shadow over the far side of the wake and therefore the laser sheet is unable to illuminate the full wake.
Due to this constraint, for the region $x/d < 6.59$, two times the temporal average of the instantaneous half wake width has been plotted in figure \ref{fig: Splitter wake width}. 
%to the instantaneous wake width that has been previously displayed in figure \ref{fig: Dynamic wake width}, since only one half of the wake may be accessed by the laser sheet.
%Due to this constraint, the metric being examined here is the mean wake coverage.

\begin{figure}
	\hspace{-0.5cm}
	\begin{tikzpicture}
	\node[anchor= west,inner sep=0] (0) at (0,1.27)
	{\includegraphics[width = 0.5\textwidth]{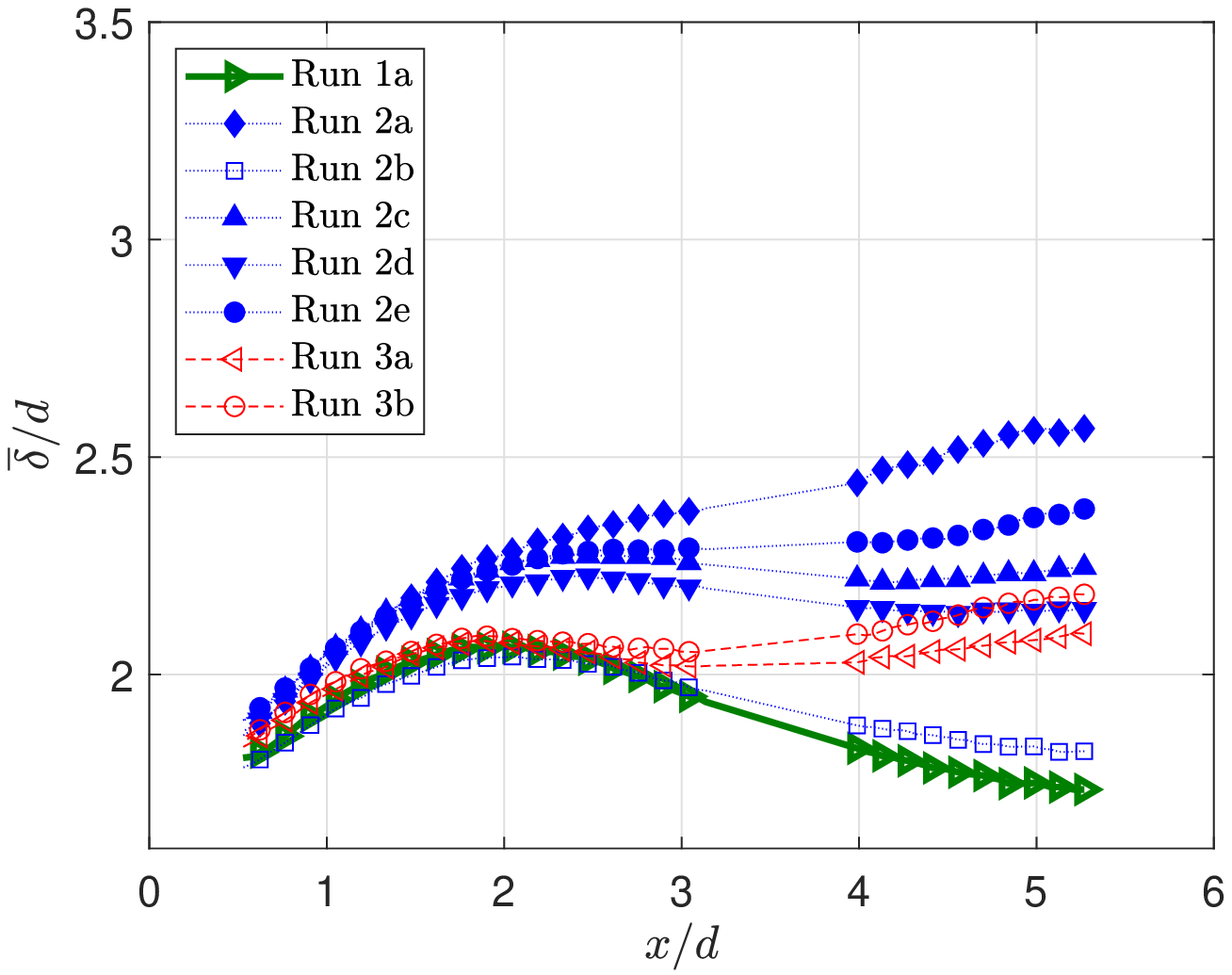}
		\includegraphics[width=0.5\textwidth]{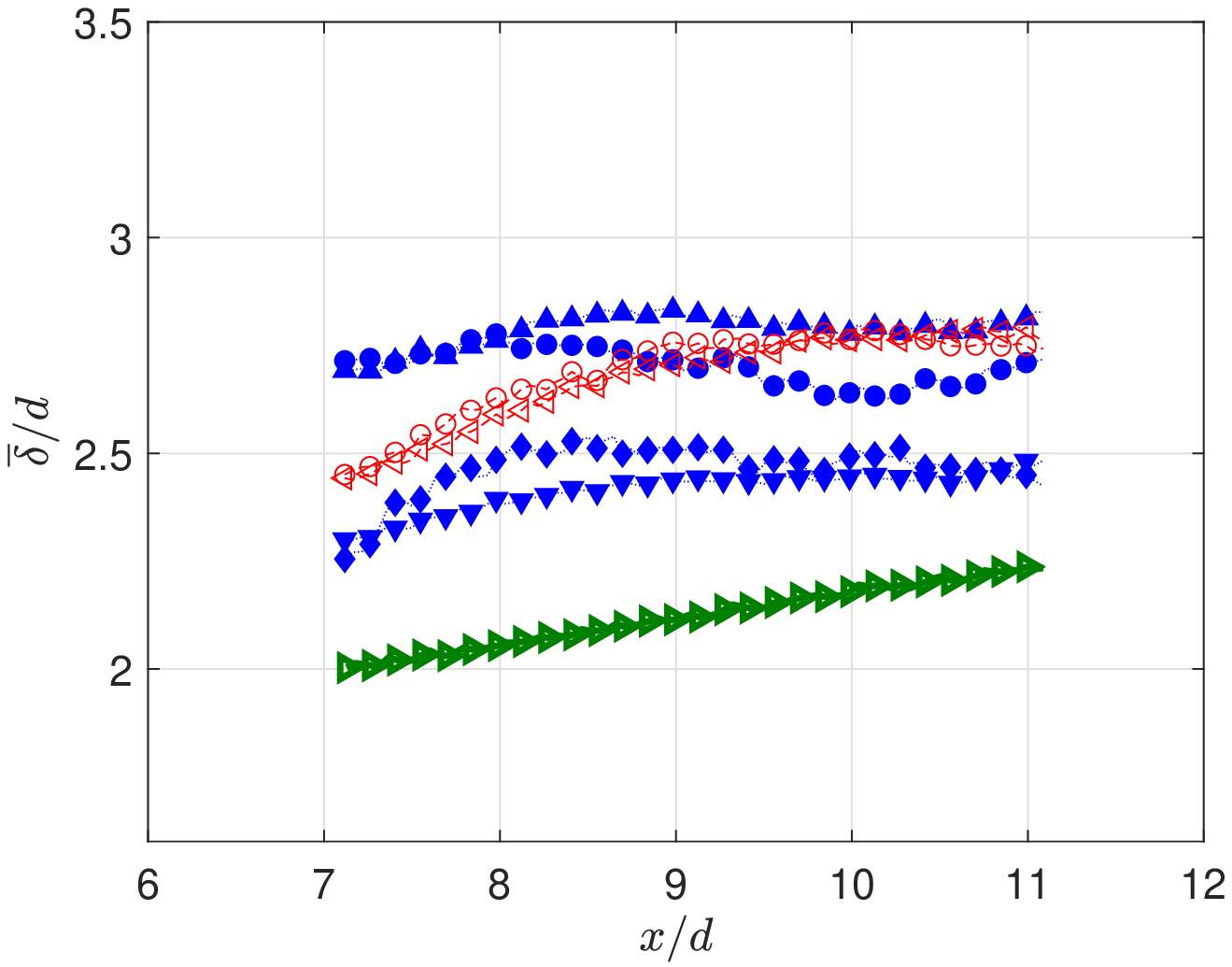}};
	\node (a) at (0,3.65) {(\textit{a})};
	\node (b) at (8.5,3.65) {(\textit{b})};
	\end{tikzpicture}
	\caption{Temporal average of the instantaneous wake width downstream of the cylinder - splitter plate combination. Plots (\textit{a}) and (\textit{b}) represent the upstream and downstream data set respectively.} 
	\label{fig: Splitter wake width}
\end{figure}

It is evident from figure \ref{fig: Splitter wake width}(\textit{a}), that the mean coverage of the wake is greatly affected by the incoming background turbulence.
Every case of subjected turbulence acts to increase the mean extent of the wake when the splitter plate is installed.
The departure in behaviour of background turbulence cases with respect to the no-grid case occurs in the region where the splitter plate is still present ($x/d < 6.59$).
Whilst, the no-grid case sees a reduction in mean width in the region of the reattachment point and further downstream on the splitter plate, similar behaviour is not observed for cases with free-stream turbulence.
The wake in these cases usually tends to continually increase in width as the $x$-position is increased, as seen in figure \ref{fig: Splitter wake width}(\textit{a}).
The behaviour in the upstream set of data is dominated by reattachment of the flow to the splitter plate.

\begin{figure}
	\centering
	\includegraphics[width = 0.6\textwidth]{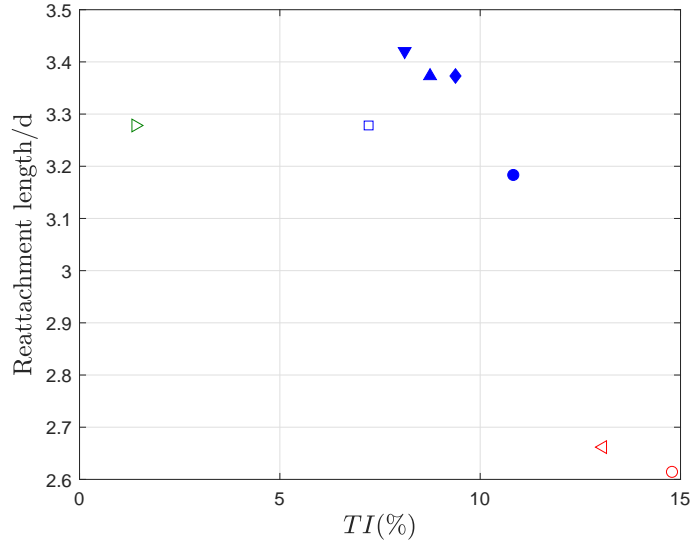}
	\caption{Reattachment length along the splitter plate, as a function of incoming turbulence intensity}
	\label{fig: Reattachment length}
\end{figure}

Freestream turbulence has a large influence on the reattachment point of the separated flow back onto the splitter plate.
\cite{Bearman1983} report on similar results for a rectangular cylinder with a splitter plate and show a decrease in reattachment length with increasing turbulence intensity.
This behaviour is recreated here for a circular cylinder, with the group 3 cases showing a significant decrease in the reattachment length (see figure \ref{fig: Reattachment length}).
In this context, the reattachment length is defined using a metric that is designed to mimic the behaviour of the wall shear-stress.
We identify the streamwise position along the splitter plate at which the wall-normal velocity gradient at the closest available $y$-position, $\dt \overline u / \dt y(y\rightarrow 0) \sim \tau_w$, changes sign.
This wall-normal position was $\lesssim 1.5$ mm above the splitter plate which corresponds to $\lesssim \theta/2$ where $\theta$ is the momentum thickness, defined in the classical way. 
The reduction in reattachment length is mainly driven by the intensity of the background turbulence rather than the length scale, which is in agreement with results reported in \cite{Bearman1983}. 
Only group 3 cases, which have a significant intensity in the background show a reduction in the reattachment length. 
Group 2 cases have a reattachment length that is largely similar to the no-grid case.

In the downstream set of data (figure \ref{fig: Splitter wake width}(\textit{b})), the wake tends to develop quickly once both halves of the wake have merged downstream of the trailing edge of the splitter plate. 
This is also the region where the early influences of the far-field effect of entrainment are visible, i.e. the suppression of nibbling \citep{Kankanwadi2020}.
Assessing the gradients of each plot $\dt \overline \delta / \dt x$, the wake in the no-grid case continues to grow at a steady rate downstream of the splitter plate.
Cases that are exposed to background turbulence, generally see an initial jump in the wake width as the two sheared regions merge. 
However, entrainment effects of the free-stream turbulence quickly act, causing a sharp decline in measured gradients across all cases with background turbulence.
This is especially evident upon applying a linear regression to points that lie between $9<x/d<11$. The no-grid case in this region continues to present a positive gradient, whereas cases in both groups 2 and 3 display a gradient close to zero.

It is clear to see that in the early developmental stages of the wake, free-stream turbulence has a very large impact and increases wake width.
It is likely that this is associated with the increased importance of entrainment through engulfment in the near-wake of a circular cylinder, further evidenced by the splitter plate experiment.
After ``switching off'' the large-scale coherent motions (and as a result meandering) with the use of the splitter plate, it was shown that background turbulence acts to increase entrainment in the region of the recirculation zone, until the sheared regions from both sides of the splitter plate interact.
Following this downstream location the wakes tend to develop quickly and the far-field effects of background turbulence on entrainment take over, suppression of nibbling leading to a slower growth rate for the wakes exposed to freestream turbulence relative to a non-turbulent background.
The analysis in this section clearly highlights the spatial influence on entrainment effects of background turbulence, an avenue of research that merits further work not least searching for the ``crossover'' point whereby the presence of freestream turbulence goes from enhancing the entrainment rate to reducing it.

\section{Identity loss events}

In the previous sections, near-field effects of freestream turbulence were assessed in a mean sense. 
The final section aims to analyse a dynamic phenomenon that has implications for instantaneous entrainment.
During the course of analysing the near-wake data, it was noticed that freestream turbulence prompted several phases of severe disruption to the wake where the regular shedding pattern and entrainment rate into the near wake were affected. 
During these intermittent events, there was a complete loss of structure in the wake and hence they are referred to as ``identity loss'' events.
These events did not occur frequently but when they did there was a significant change to the structure, and dynamics, of the wake.
The temporary loss of the large-scale von K\'{a}rm\'{a}n vortices meant that the wake meandering was completely suppressed for the duration of the identity loss event.
We further postulate that they reduce the instantaneous entrainment rate into the wake since the absence of the large-scale coherent motions is likely to reduce large-scale engulfment substantially.
Due to their relative scarcity it was sadly not possible to quantify their effect on the mean entrainment/wake growth but nevertheless we report on their phenomenology.

\begin{figure}
	\centering
	\begin{tikzpicture}
	\node[anchor= west,inner sep=0] (0) at (0,1.27) {\includegraphics[width=\textwidth]{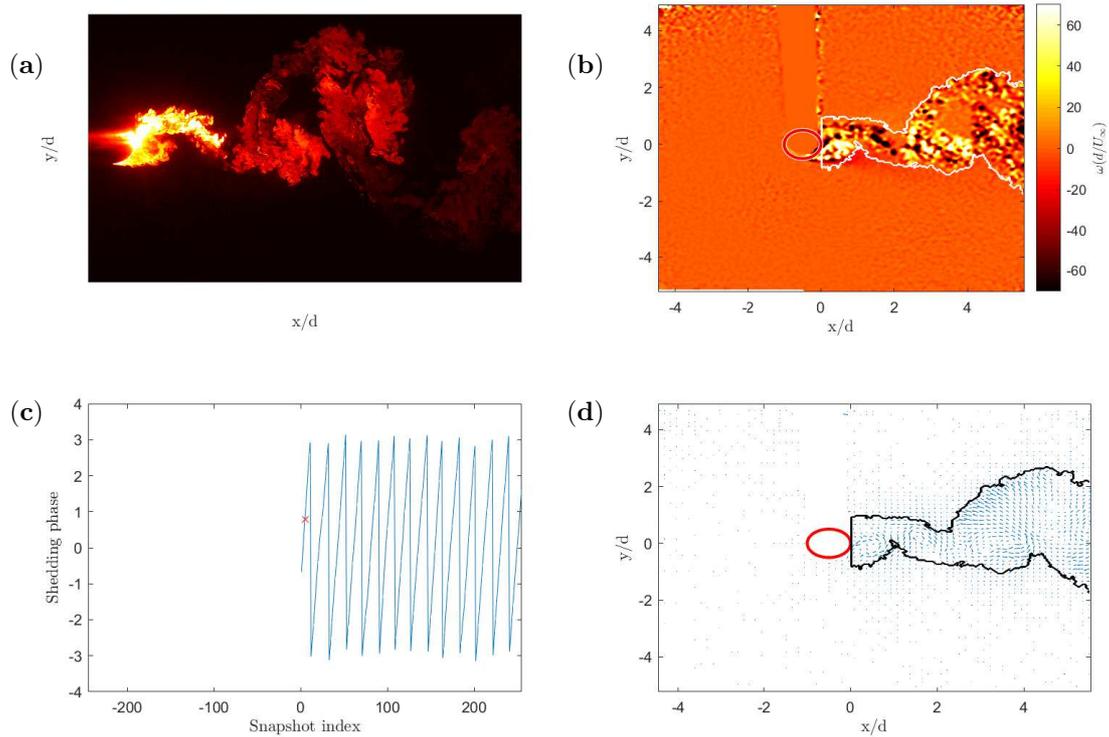}};
	\node (0) at (1.4,5.2) {(\textbf{a})};
	\node (1) at (8.75,5.2) {(\textbf{b})};
	\node (2) at (1.4,0.6) {(\textbf{c})};
	\node (3) at (8.75,0.6) {(\textbf{d})};
	\end{tikzpicture}
	\caption{Regular shedding for a cylinder subjected to an undisturbed free-stream. The figures show details of a single investigated snapshot. (\textbf{a}) Raw PLIF image of the investigated snapshot. (\textbf{b}) Normalised vorticity, $ \omega_z/(U_{\infty}/d) $, is plotted as the background ($ U_{\infty} $, is the free-stream velocity and  $ d $, is the cylinder diameter). Note the solid white contour line represents the scalar interface. (\textbf{c}) Phase signal for the reconstructed shedding mode, obtained using optimal mode decomposition \citep{Wynn2013}. The red cross represents the snapshot being examined in this figure. (\textbf{d}) Instantaneous fluctuating velocity. Note the overlayed contour represents the scalar interface.}
	\label{fig: regular shedding}
\end{figure}

\begin{figure}
	\centering
	\begin{tikzpicture}
	\node[anchor= west,inner sep=0] (0) at (0,1.27) {\includegraphics[width=\textwidth]{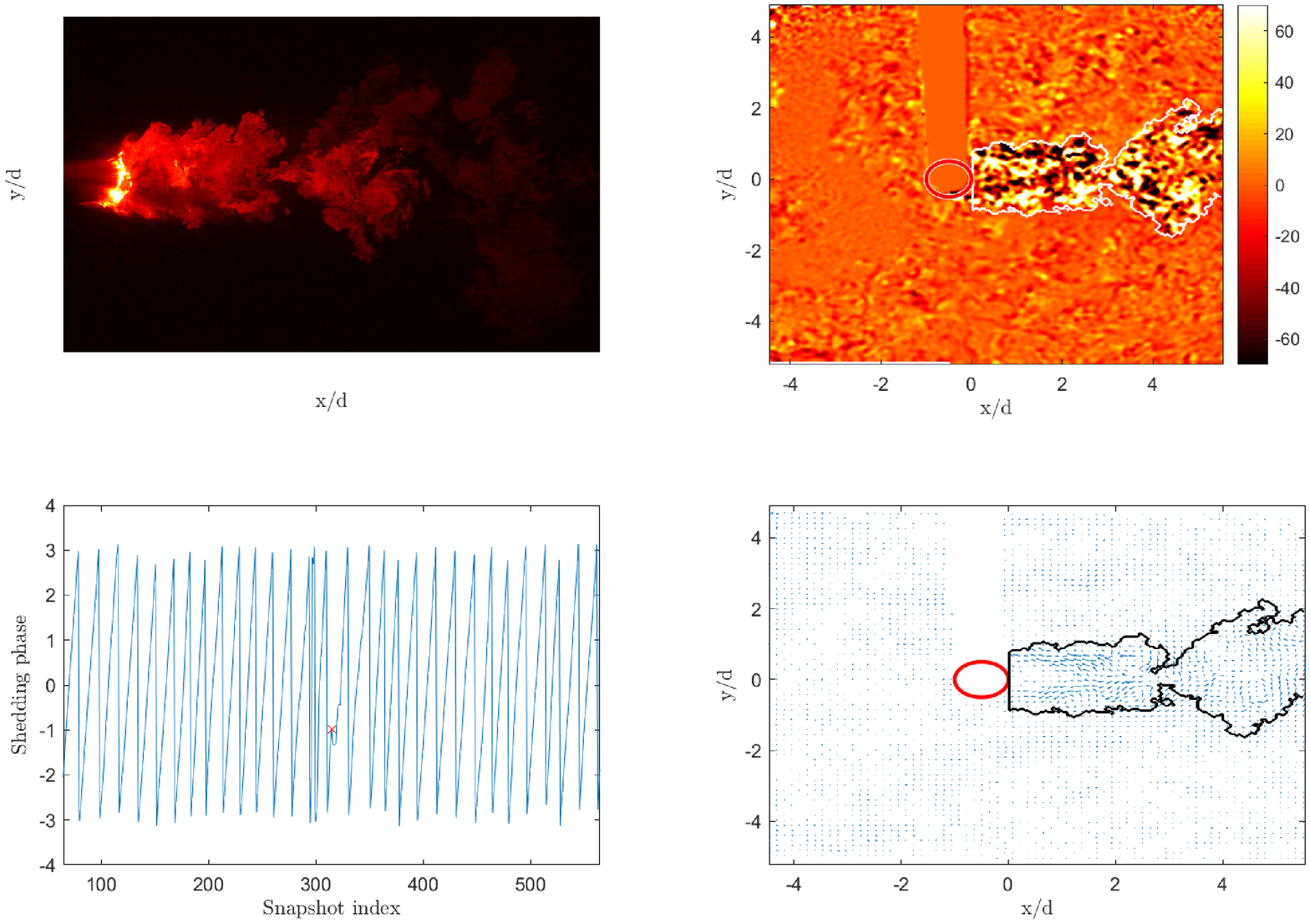}};
	\node (0) at (1.4,5.2) {(\textbf{a})};
	\node (1) at (8.75,5.2) {(\textbf{b})};
	\node (2) at (1.4,0.6) {(\textbf{c})};
	\node (3) at (8.75,0.6) {(\textbf{d})};
	\end{tikzpicture}
	\caption{A snapshot of an Identity loss event, an example of altered shedding experienced by a cylinder in the presence of free-stream turbulence The figures show details of a single investigated snapshot. (\textbf{a}) Raw PLIF image of the investigated snapshot. (\textbf{b}) Normalised vorticity, $ \omega_z/(U_{\infty}/d) $, is plotted as the background ($ U_{\infty} $, is the free-stream velocity and  $ d $, is the cylinder diameter). Note the solid white contour line represents the scalar interface. (\textbf{c}) Phase signal for the reconstructed shedding mode, obtained using optimal mode decomposition \citep{Wynn2013}. The red cross represents the snapshot being examined in this figure. (\textbf{d}) Instantaneous fluctuating velocity. Note the overlayed contour represents the scalar interface.}
	\label{fig: altered shedding}
\end{figure}

Figures \ref{fig: regular shedding} and \ref{fig: altered shedding}, respectively, highlight the near wake during regular shedding in an undisturbed background and during an altered shedding event noticed under the influence of freestream turbulence.
These events where the shedding mechanism is seemingly altered are identifiable as ``phase hiccups'' in the phase signal of the shedding mode.
Recall that the phase signal is obtained by evaluating the angle of the time dependent complex coefficients required to reconstruct the shedding mode in the optimal mode decomposition (OMD) technique (see \S \ref{sec: Triple decomposition}).
Figures \ref{fig: regular shedding}(\textit{c}) and \ref{fig: altered shedding}(\textit{c}) highlight the OMD phase signal during each event.
The departure from periodic fluctuation of the phase is evident during an altered shedding event.
Therefore, the OMD phase signal was used to identify occurrences of these events within all the runs conducted in this campaign.

Rather than periodically shed vortices as observed in a von K\'arm\'an vortex street, an incoherent shear layer is shed as the wake of the cylinder during an identity loss event. 
This in-turn gives rise to Kelvin-Helmholtz type vortices in both shear layers either side of the centreline. 
As can be observed in figure \ref{fig: altered shedding}(\textit{d}), the instabilities develop into a counter-rotating pair of vortices and cause significant back-flow near the wake centreline.
The implications with regard to instantaneous entrainment rate are likely to be significant. 
Furthermore, it is highly likely that a sudden intermittent change to the shedding mechanism of the cylinder may have consequences with respect to the instantaneous load placed on the cylinder.

\begin{figure}
	\hspace{-0.5cm}
	\begin{tikzpicture}
	\node[anchor= west,inner sep=0] (0) at (0,1.27)
	{\includegraphics[width = 0.5\textwidth]{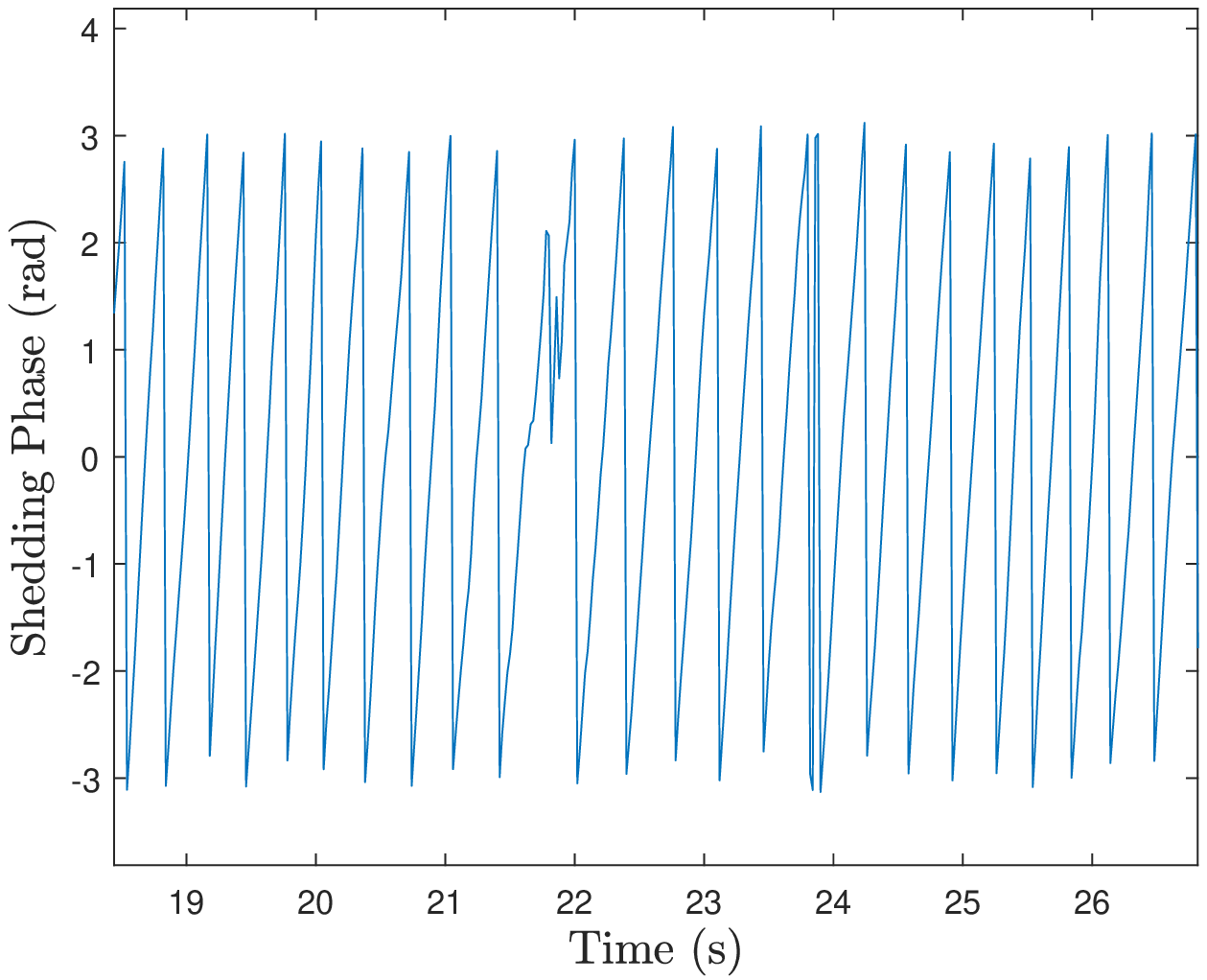}
		\includegraphics[width=0.5\textwidth]{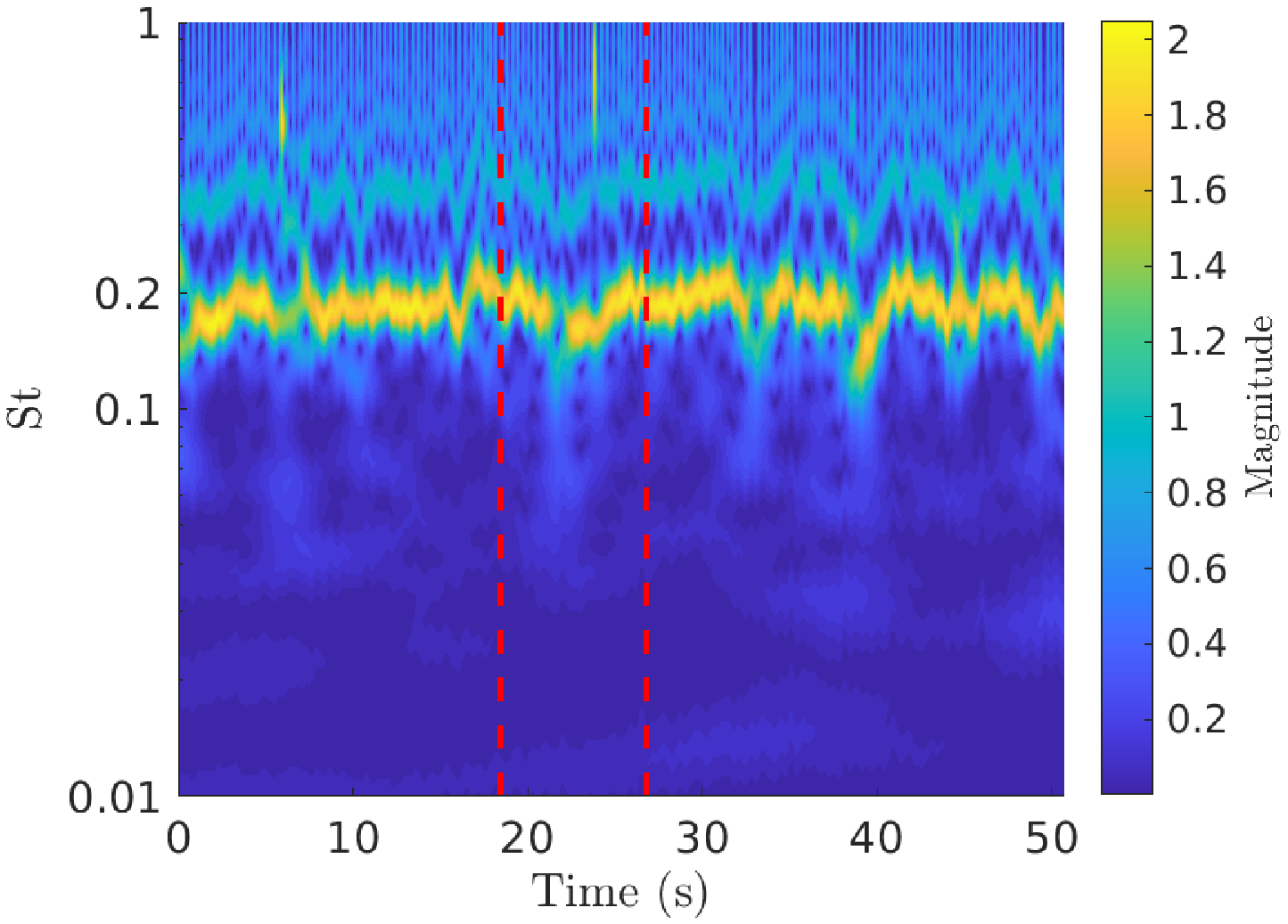}};
		\node[anchor= west,inner sep=0] (1) at (0,-6)
		{\includegraphics[width=0.5\textwidth]{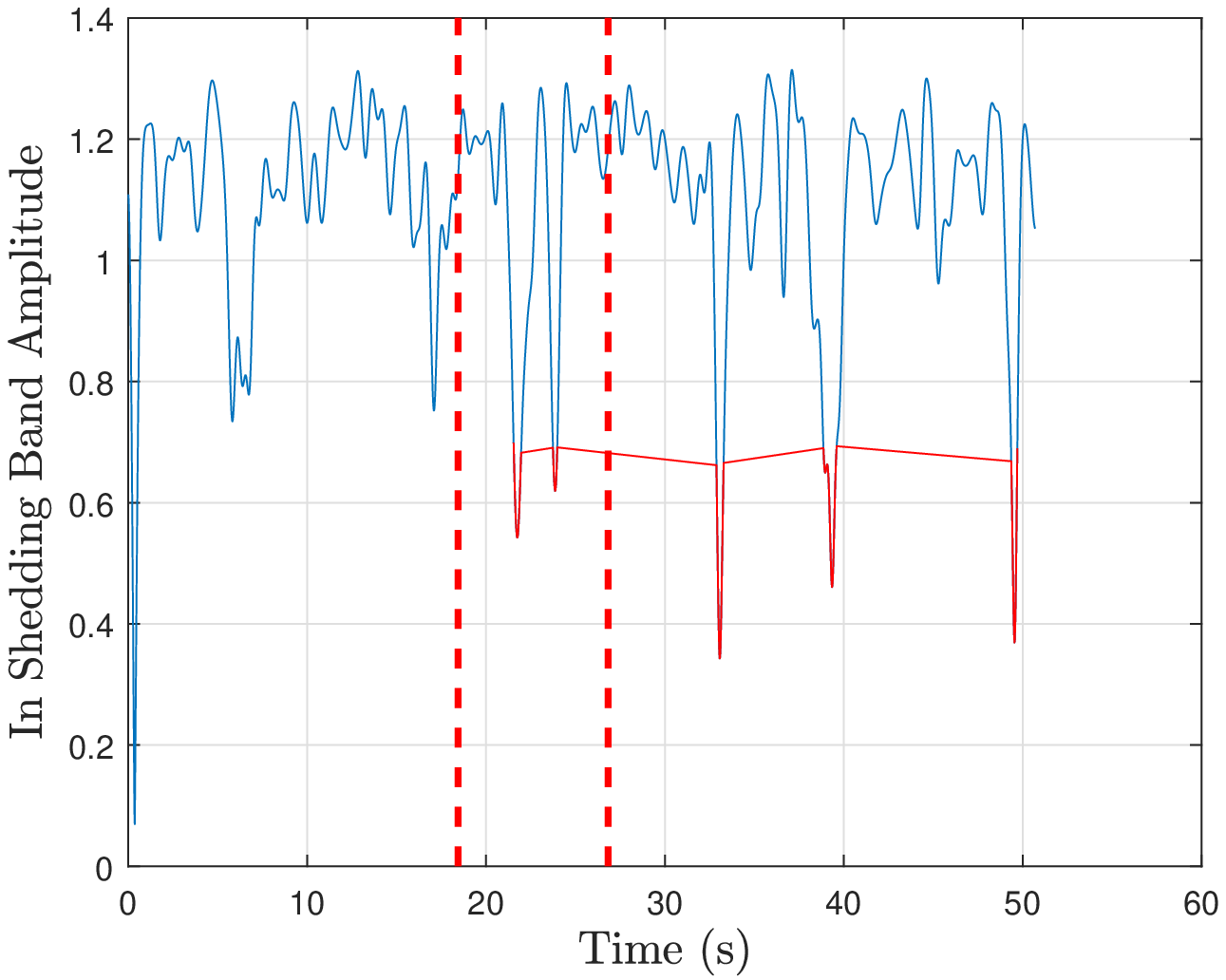}
    		\includegraphics[width=0.5\textwidth]{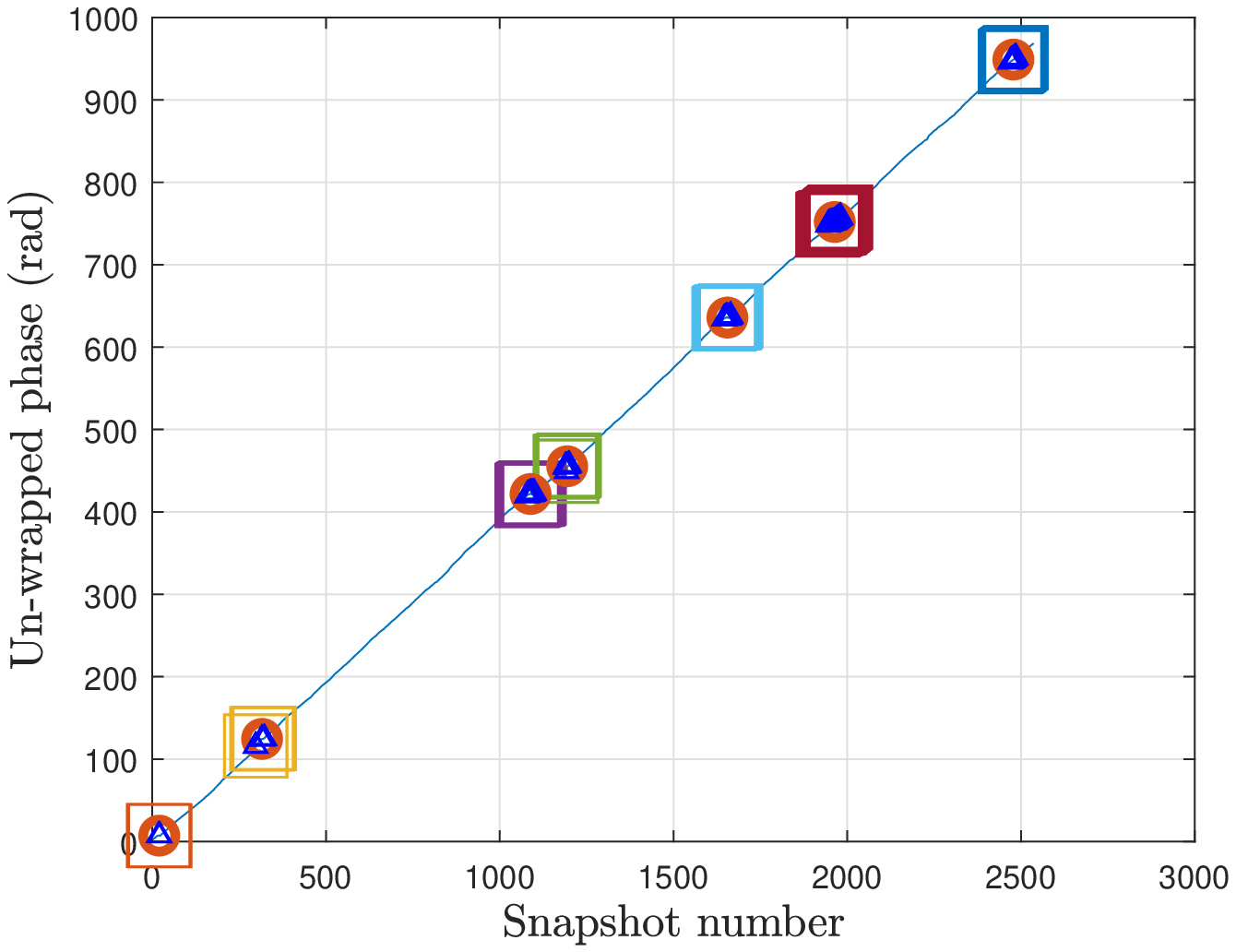}};
	\node (a) at (4.2,-2.35) {(\textit{a})};
	\node (b) at (12.7,-2.35) {(\textit{b})};
	\node (c) at (4.2,-9.7) {(\textit{c})};
	\node (d) at (12.7,-9.7) {(\textit{d})};
	%\node (a) at (0.5,3.5) {(\textit{a})};
	%\node (b) at (9,3.5) {(\textit{b})};
	%\node (c) at (0.5,3.5-6.27) {(\textit{c})};
	%\node (d) at (9,3.5-6.27) {(\textit{d})};
	\end{tikzpicture}
	\caption{Plots describing the methodology to identify ID loss events. (\textit{a}) Example of the phase signature of the OMD shedding mode during an ID loss event. (\textit{b}) Magnitude scalogram of the continuous wavelet transform of the phase signal. (\textit{c}) Amplitude within the shedding band as a function of time. The parts highlighted in red indicate regions of shifted frequency. The red dotted line in (\textit{b}) and (\textit{c}) outline the time window displayed in the top-left plot. (\textit{d}) Clusters of identified ID loss events are superimposed on top of the cumulative phase.} 
	\label{fig: ID Loss methodology}
\end{figure}

The plots in figure \ref{fig: ID Loss methodology} depict the methodology employed to automate the identification of these events. 
Note that these plots are typical examples taken from Run 2e.
Figure \ref{fig: ID Loss methodology}(\textit{a}) is a plot of the typical phase signature of an altered shedding event. 
Reversal of the phase signal as well as a significant shift in instantaneous shedding frequency are characteristic indications of underlying identity loss events.
Phase reversals were identified by scanning for a change in gradient over the time series after accounting for the discontinuities in phase at the end of each cycle.
This technique was combined with spectral analysis techniques, described below, to allow for the identification of frequency-change events.
To this end the phase signal was analysed using a continuous wavelet transform.
Figure \ref{fig: ID Loss methodology}(\textit{b}) depicts the magnitude scalogram of the shedding phase transformed using a Morse wavelet.
Clear departures in the magnitude at the shedding frequency are evident in the figure.
The amplitude within the shedding band can be extracted and is plotted in figure \ref{fig: ID Loss methodology}(\textit{c}).
A user defined threshold may then be used to classify snapshots where the in-band amplitude drops below the threshold as part of an identity loss event. 
In this case, an amplitude threshold of 0.7 was selected.
Snapshots that passed either of the two tests, phase reversal or frequency shift, were deemed to be of interest. 
These snapshots were then clustered into distinctive events using a subtractive clustering methodology.
Clusters of snapshots are depicted in figure \ref{fig: ID Loss methodology}(\textit{d}) where they are superimposed over the unwrapped (cumulative) phase signal giving an indication of their intermittency.

Further investigation into these altered shedding events highlights the similarity to the shear layer instability phenomena, which develops through the action of a Kelvin-Helmholtz mechanism. \cite{Prasad1997} conduct a detailed investigation of this phenomenon in the near-wake of a cylinder subjected to laminar flow. The vortices that form in the separating shear layer generally scale with the thickness of the shear layer and are visually similar to those found in a plane mixing layer between two streams of differing velocity. These similarities stretch across to phenomena observed in this study. \cite{Prasad1997} show that shear layer development is significantly influenced by three-dimensional near-wake phenomena such as parallel and oblique shedding conditions.
Note that these conditions refer to vortex tubes being shed either parallel to the cylinder axis or at an oblique angle to the cylinder axis respectively. Typically the oblique shedding angles are between $15\operatorname{-}20^{\circ}$ \citep{Williamson1996}.
\cite{Prasad1997} control the shedding conditions by forcing them into the required condition with the use of inclined endplates. A clear peak is visible at the shear layer frequency for parallel conditions, whereas, it is suggested that the instability is moderated by oblique shedding conditions.
Akin to the altered shedding events observed in this study, the shear layer instability is also inherently intermittent.
\cite{Prasad1997} show that intermittency increases with both Reynolds number and downstream distance of the probe.
Furthermore, they highlight the distinct possibility of spontaneous generation of three-dimensional structures akin to structures referred to as mode B by \cite{Williamson1996}, to contribute significantly to the intermittency.
Mode B structures indicate three-dimensional streamwise structures that have a typical span-wise wavelength of 1 diameter.

\begin{figure}
	\hspace{-0.5cm}
	\begin{tikzpicture}
	\node[anchor= west,inner sep=0] (0) at (0,1.27)
	{\includegraphics[width = 0.5\textwidth]{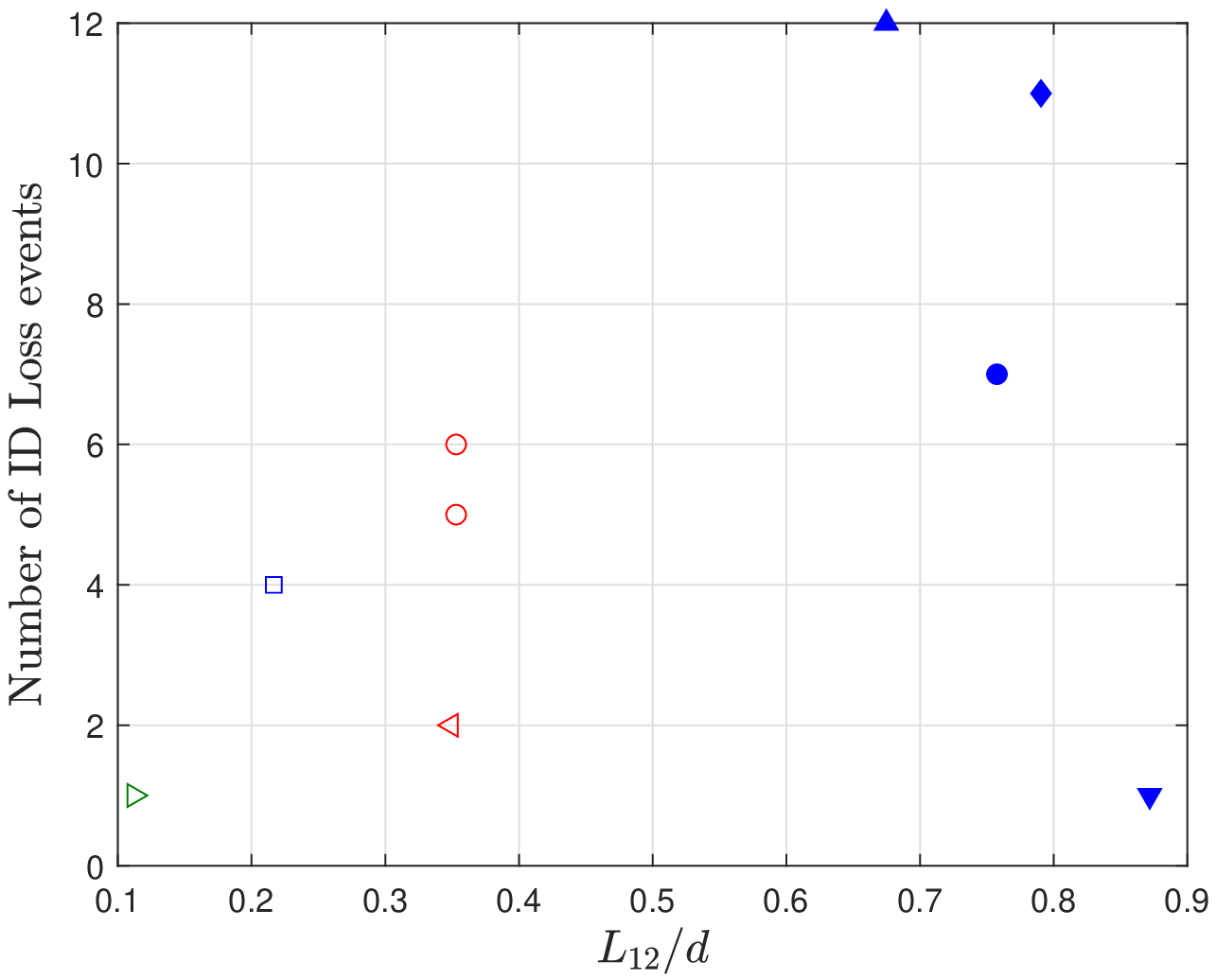}
		\includegraphics[width=0.5\textwidth]{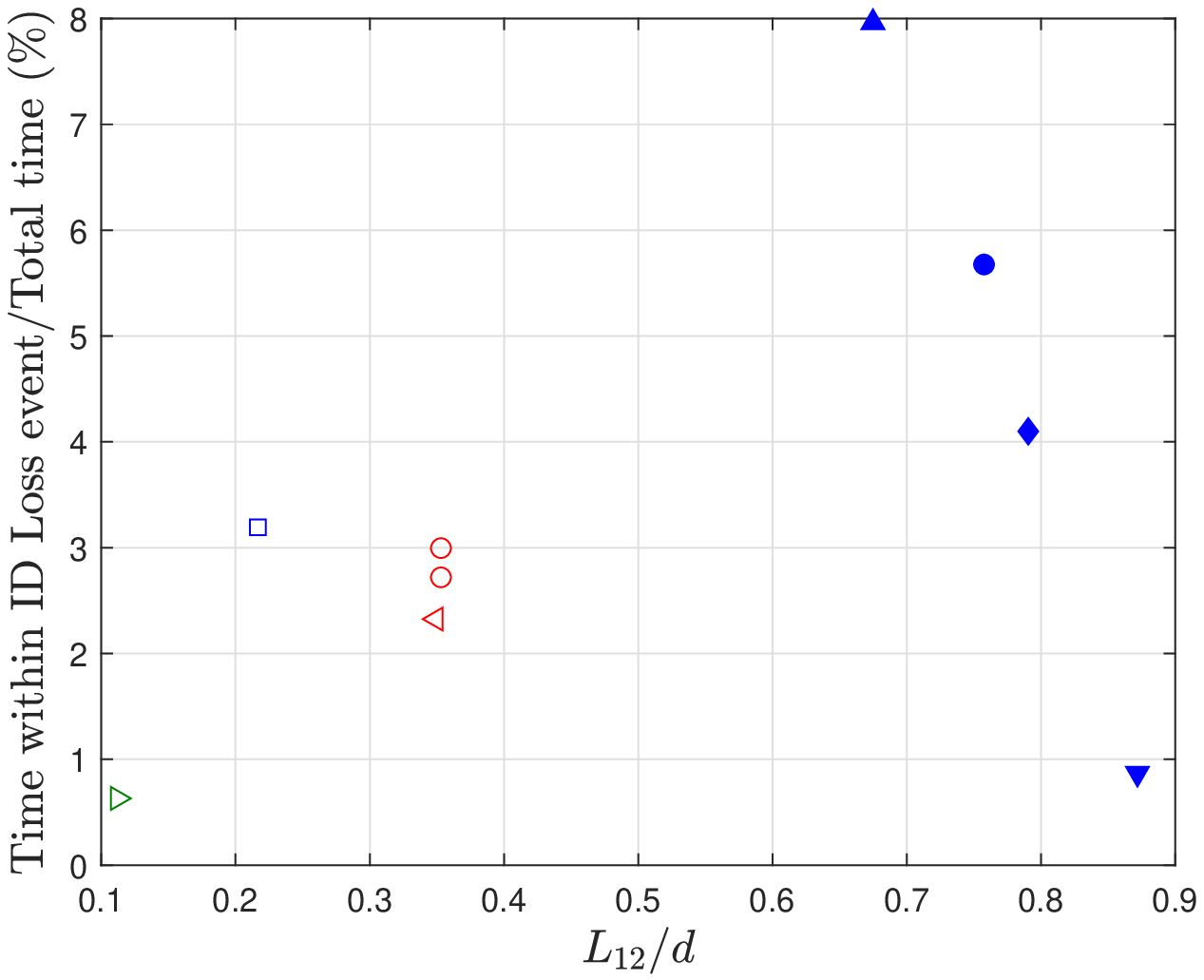}};
	\node (a) at (0,3.65) {(\textit{a})};
	\node (b) at (8.5,3.65) {(\textit{b})};
	\end{tikzpicture}
	\caption{Intermittency as a function of subjected integral length scale. (\textit{a}) Number of events where the instability occurred. (\textit{b}) Percentage of time spent inside an ID loss event during the run.}
	\label{fig: ID Loss stats}
\end{figure}

The intermittency of the instability observed in this study is investigated in figure \ref{fig: ID Loss stats}. Figure \ref{fig: ID Loss stats}(\textit{a}) depicts the number of identified events (clusters) during a run. Figure \ref{fig: ID Loss stats}(\textit{b}), on the other hand aims to quantify intermittency as a percentage of the time period that is spent within identity loss events with respect to the total time.
Whilst no clear trend with respect to subjected turbulence intensity is to be seen, the incoming integral length scale has some influence.
Both, the number of observed events as well as the fraction of time when instability events are present, generally increase as a function of integral length scale.
Run 2d continues to be an outlier with significantly reduced manifestation of identity loss events. %\textcolor{red}{The square fractal grid in this case is placed approximately 3.5 m upstream of the field of view and safely in the decay region of turbulence production. This significantly increased distance meaning that the turbulence is in decay may have a reduced influence to near-wake behaviour.}
It is possible that free-stream turbulence conditions with length scales of the order of 1 cylinder diameter cause an excitation of the instability similar to the previously discussed mode B structures.

\begin{figure}
	\hspace{-0.5cm}
	\begin{tikzpicture}
	\node[anchor= west,inner sep=0] (0) at (0,1.27)
	{\includegraphics[trim=3cm 0 3cm 0,clip,width = 0.5\textwidth]{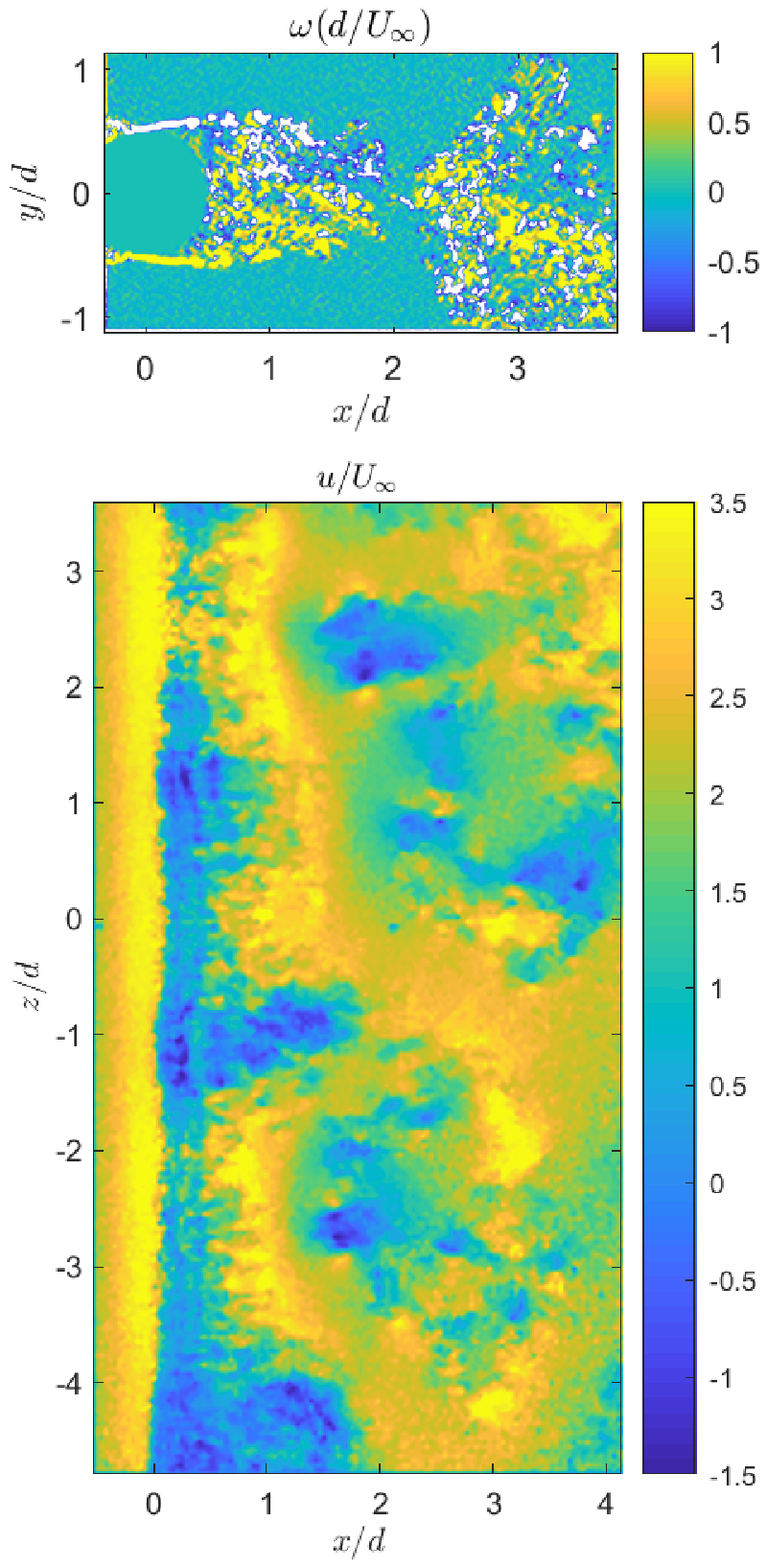}
		\includegraphics[trim=3cm 0 3cm 0,clip,width=0.5\textwidth]{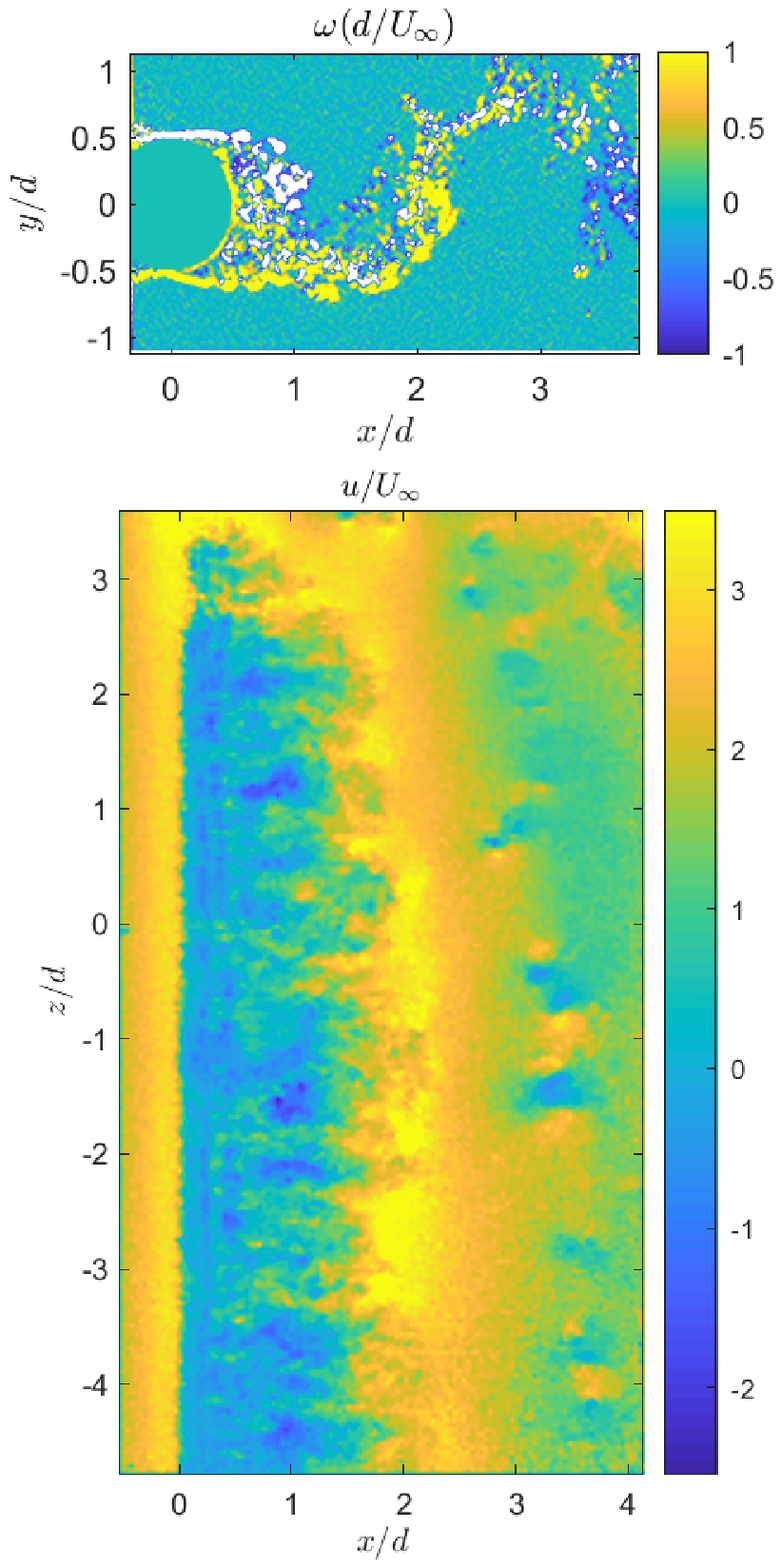}};
	\node (a) at (4.2,-7.5cm) {(\textit{a}) Altered shedding event};
	\node (b) at (13,-7.5cm) {(\textit{b}) Regular shedding};
	\end{tikzpicture}
	\caption{Spanwise flow structures during (\textit{a}) an altered shedding event and (\textit{b}) regular shedding. Note, the $x-z$ plane is located at $y=-0.5d$. Data obtained from \cite{Cicolin2020Thesis}}
	\label{fig: Spanwise ID loss}
\end{figure}

Finally, we comment on three-dimensional flow structures downstream of the cylinder during the instability by utilising data collected by \cite{Cicolin2020Thesis}. 
They investigated the flow downstream of cylinder - control rod configurations, through planar PIV experiments on dual planes that are orthogonal to each other, allowing the collection of $u$, $v$ and $w$ velocities. 
Identity loss events can also be observed in their data-set. 
OMD analysis can be used on the data collected in the $x-y$ plane to identify instances of the altered shedding events. 
Data from the $x-z$ plane can then be used to comment on any spanwise structures.
Figure \ref{fig: Spanwise ID loss}(\textit{a}) depicts the flow field during an identified altered shedding event. 
Dislocations in the spanwise component are clear to see, as it appears that the shedding mechanism is neither uniquely oblique or parallel during the event.
On the other hand, \ref{fig: Spanwise ID loss}(\textit{b}) indicates the flow field during regular parallel shedding, where no dislocations are visible.
This adds evidence to the suggestion that manifestations of identity loss events are inherently controlled by three-dimensional structures.
It is therefore plausible to reason that identity loss events are excited by the presence of length scale matched background turbulence as they provide an increased content of three-dimensional structures that impinge on the cylinder.

Symmetrical vortex shedding behaviour that is observed during identity loss events is also very similar to the behaviour noticed by \cite{Barbi1986} and \cite{Konstantinidis2007} during their investigations of vortex shedding of a circular cylinder when subjected to oscillatory flow. 
\cite{Barbi1986} highlight the importance of lock-on behaviour by the subjected oscillatory flow that has a tendency of attracting the shedding frequency towards the driving frequency of the background for frequency ratios that are under the lock-on threshold.
Most importantly to the current study, they show that for frequency ratios, $F_d/F_{s0} = 1$ (where $F_d$ is the driving frequency and $F_{s0}$ is the shedding frequency in steady conditions), two symmetric vortices are shed simultaneously for the cylinder in one cycle.
\cite{Konstantinidis2007} specifically study this symmetrical shedding behaviour at a range of forcing frequencies. 
They show that symmetrical shedding is not stable and gives way to anti-symmetric shedding further downstream and also that the initial formation of the symmetric vortices on both sides of the cylinder is always phase locked to the subjected flow, thereby meaning that this behaviour is a direct consequence of the subjected symmetric perturbation.
The subjected freestream turbulence in this study is spectrally rich and it is possible that the upstream flow conditions may be conducive to conditions required to produce the symmetrical shedding mode.
It is however, imperative to highlight that \cite{Konstantinidis2007} lean on the shear layer instability behaviour outlined by \cite{Prasad1997} and state similarities between their two cases.
The forcing flow (in their case) or the subjected free-stream turbulence (in this study) interacts with the separated shear layers to produce the observed phenomena.

The shear layer instability phenomena reported by \cite{Prasad1997} was observed to be on the whole, two-dimensional along the span and in-phase across the wake.
However, those experiments were conducted in a non-turbulent free-stream.
Conditions in this study are unlikely to recreate the two-dimensionality observed by \cite{Prasad1997} due to the turbulent and dynamic nature of the flow subjected to the circular cylinder.
\cite{Norberg2003} provides a potential alternative explanation, should the altered shedding events not be two-dimensional along the span.
They identify two shedding modes that co-exist within the sub-critical regime; namely, the ``high quality" and ``low quality" modes. 
They are named as such due to the spectral quality of the shedding frequency within each of the two modes.
They explain that the \textit{high quality} mode yields fairly regular vortex shedding with only minor spanwise undulations of von K\'{a}rm\'{a}n shedding.
Whereas, the \textit{low quality} mode may be considered to be similar to identity loss events observed in this study.
This mode is primarily governed by a Kelvin Helmholtz instability mechanism causing disruptions and random forcing to the vortex shedding process.
Lastly, \cite{Norberg2003} states that these disruptions lead to randomly positioned vortex dislocations along the span.

To summarise, the action of the incoming free-stream turbulence manifests itself through the formation of altered shedding events named as identity loss events due to the loss of similarity to regular vortex shedding. There are observed similarities to the shear layer instability phenomena and the action of incoming turbulence is seen to encourage the manifestation of these events. More specifically, cases with higher length scale in the background turbulence are seen to increase the frequency at which the instability occurs. The intermittency of this behaviour is likely to be explained by the three-dimensional perturbations of the incoming flow. Structures with length scales that are similar to the diameter of the cylinder are likely to excite the observed phenomena.
Finally, a couple of alternative explanations are also laid out for the reader.

\section{Conclusions}

The near-field development of a circular cylinder wake subjected to  various ``flavours'' of freestream turbulence, in which the integral length scale and turbulence intensity was independently varied, was experimentally examined using simultaneous PIV and PLIF experiments. 
A control experiment in which no turbulence-generating grid was mounted upstream of the cylinder was also conducted.
The oft-reported conclusion of increased near-wake width in the presence of free-stream turbulence, in comparison to a non-turbulent background, was recovered when the time-averaged spatial extent of the instantaneous wake position was considered.
This result is why wind-turbine wake models typically consider an enhanced wake spreading rate as turbulence intensity is increased, \citep[e.g.][]{Niayifar2016}.
However, this metric amalgamates, in effect, two different phenomena: the growth rate of the wake instantaneously due to entrainment of background fluid and the envelope of the spatial extent of the ``meandering'' of the wake due to the presence of large-scale von K\'{a}rm\'{a}n vortices embedded within the wake.
Whilst wind turbine wakes are different from the cylinder wakes considered in the current work they also experience meandering \citep{Okulov2014} and contain multiple large-scale vortical structures such as helicoidal tip vortices \citep{Vermeer2003}.
It was shown that the integral length scale of the freestream turbulence had a defining impact on the behaviour of near-field wake development. 
Cases with large integral length scales experienced a reduction in instantaneous wake width but a substantially increased amplitude of wake meandering.
The combination of these results therefore still led to an increased wake width in comparison to the control, no-grid case.
Cases that had a smaller integral length scales in the freestream turbulence experienced an increase in both entrainment rate and an increase in the amplitude of meandering. 
This, therefore, again resulted in an overall increase of the time-averaged wake width.
We further showed that the turbulence within the wake remains distinct from the background turbulence, since the velocity fluctuations de-correlate over a distance substantially shorter than the wake half width.
Nevertheless, background turbulence with a larger characteristic length scale seems to induce larger turbulent eddies within the wake itself.

Purely from an entrainment perspective, it was observed that cases with the highest turbulence intensity within the freestream turbulence led to the largest entrainment rates in a time-averaged sense.
This result is in contrast to that of \citet{Kankanwadi2020} who showed that in the far field of a cylinder-wake, turbulence intensity was inversely correlated to entrainment rate, even leading to net detrainment for some cases where the turbulence intensity of the background was greater than that within the wake.
We therefore postulate that the effect of freestream turbulence has differing effects on the two different mechanisms of turbulent entrainment, engulfment and nibbling.
Nibbling is primarily driven by small-scale diffusion and this was observed to be reduced in the far field by \citet{Kankanwadi2020} since freestream turbulence had the effect of promoting intermittent, powerful detrainment events, thereby reducing the net entrainment rate.
We now postulate that freestream turbulence enhances large-scale engulfment of background fluid.
Since engulfment is the dominant entrainment mechanism in the near field of turbulent shear flows in which large-scale coherent structures are active \citep{Yule1978} we postulate that the interaction between background turbulence and large-scale coherent structures acts to enhance the turbulent entrainment rate in the near field.
Such a postulation therefore predicts the existence of a ``crossover'' point at some streamwise position where the coherent structures have sufficiently decayed such that the net negative effect of suppression of nibbling overtakes the enhancement of engulfment provided by freestream turbulence and the net entrainment rate switches from being enhanced to suppressed in comparison to a non-turbulent background.
Identifying this ``crossover point'' is a key focus of our future work.

Further evidence for this theory was provided by a complementary experiment in which a splitter plate was attached to the rear face of the cylinder.
The objective of this experiment was to eliminate the large-scale von K\'{a}rm\'{a}n vortices in the near wake for comparison against the classical circular cylinder case.
However, the shift in the effect of background turbulence on entrainment was clear to see as the wake continued to develop. 
Wake growth was stunted for $x/d > 8$ when background turbulence was introduced. 
This was not the case for the no-grid case, as the wake in this case continued to grow steadily.
This is indicative of a reduced entrainment rate into the flow in response to the elimination of the large-scale coherent structures, i.e. evidence for the ``crossover point'' alluded to earlier.
Additionally, the splitter plate experiment showed that an increase in instantaneous wake width was observed in the immediate vicinity of the cylinder even after vortex shedding was controlled in the wake. 
In this region of the flow there is strong coherence associated to the mean re-circulation.

Intermittent events in which the regular vortex shedding was interrupted in response to the presence of freestream turbulence were identified and termed as identity loss events.
Similarities with various shear-layer-instability phenomena were explored. 
It was found that cases with higher length scale in the background turbulence increased the frequency at which these events occurred. 
Finally, a potential excitation of this phenomenon arising from encounters with freestream-turbulence structures with comparable length scales to the diameter of the cylinder was discussed.
The combined results in this manuscript show that the length scale of the freestream turbulence is arguably a more important parameter than the turbulence intensity in the near wake; a result that is in contrast to the far wake in which turbulence intensity was observed to be the dominant parameter of the freestream turbulence \citep{Kankanwadi2020}.

\appendix

\section{Optimal mode decomposition based approach to extracting the phase signal} \label{sec:omd}

OMD is a snapshot based technique that exploits a linear dependency between successive snapshots. 
A time-invariant matrix $ \textbf{D} $, of size, $ \textbf{D} \in \mathbb{R}^{m \times m}$, determines the system dynamics that transform $ \textbf{q}^{n} $ to $ \textbf{q}^{n+1} $, where $ \textbf{q} $ is a collection of 2D PIV snapshots reshaped into a 1D vector ($ \textbf{q} \in \mathbb{R}^{m \times 1}$). 
Note that $ m $ depicts the snapshot size. The system is said to be represented appropriately when \eqref{eq: DMD minimise} is minimised.

\begin{equation}\label{eq: DMD linear}
\textbf{q}^{n+1} \simeq \textbf{D} \cdot \textbf{q}^{n}
\end{equation}
\begin{equation}\label{eq: DMD minimise}
\min_{\textbf{D}}||[\textbf{q}_2,\dots,\textbf{q}_{N}] - \textbf{D} \cdot [\textbf{q}_1,\dots,\textbf{q}_{N-1}]||^2
\end{equation}

\noindent However, the difficulty of finding a solution to \eqref{eq: DMD minimise} arises due to the number of unknowns present in the problem, $ m^2 $. 
OMD aims to alleviate this problem by approximating $ \textbf{D} $ using a lower order matrix of arbitrary rank, $ r $, represented by $ \textbf{L} \cdot \textbf{M} \cdot \textbf{L}^T $. 
The problem statement is then described by \eqref{eq: OMD minimise}. 
The matrix $ \textbf{L} $ acts as a low dimensional subspace of the flow field, whereas $ \textbf{M} $ contains the dynamics of the system.

\begin{equation}\label{eq: OMD minimise}
\begin{aligned}
\min_{\textbf{L},\textbf{M}} || [\textbf{q}_2,\dots,\textbf{q}_{N}] - \textbf{L} \cdot \textbf{M} \cdot \textbf{L}^T \cdot [\textbf{q}_1,\dots,\textbf{q}_{N-1}] ||^2,\\
\textbf{L} \in \mathbb{R}^{m \times r},\ \textbf{M} \in \mathbb{R}^{r \times r},\ \textbf{L} \cdot \textbf{L}^T = \textbf{I}.
\end{aligned}
\end{equation}

\noindent Upon the optimisation of the independent variables, $ \textbf{L} $ and $ \textbf{M} $, the OMD eigenvalues, $ \lambda_n^{OMD} $, and modes, $ \Phi_n^{OMD}(\textbf{x}) $ can be evaluated as per \eqref{eq: OMD eigs} and \eqref{eq: OMD modes}

\begin{equation}\label{eq: OMD eigs}
\lambda_n^{OMD} = \frac{\ln\lambda_n(\textbf{M})}{\Delta t}
\end{equation}
\noindent where, $ \Delta t = \frac{1}{f_{aq}}$, ie. the time difference between consecutive velocity fields. 
The OMD modes can then be calculated as follows
\begin{equation}\label{eq: OMD modes}
\Phi_n^{OMD}(\textbf{x}) = \textbf{L} \textbf{z}_n
\end{equation}
Note that, $ \textbf{z}_n $, is the eigenvector corresponding to $ \lambda_n(\textbf{M}) $.
\noindent The flow field, $ s'(\textbf{x}, t) $ can then be reconstructed using the following relationship. 

\begin{equation}\label{eq: OMD reconstruction}
s'(\textbf{x}, t) \simeq \sum_{n=1}^{r} c_n \Phi_n^{OMD}(\textbf{x}) e^{\lambda_n^{OMD} t}
\end{equation}
Coefficients, $ c_n $, are required as the OMD modes are normalised to follow, $ ||\Phi_n^{OMD}||^2 = 1 $.

The triple decomposition method outlined by \cite{Baj2015}, and also the one used in this study, uses the optimal mode decomposition technique to evaluate OMD modes that represent the shedding of the cylinder. 
Shedding mode identification is done by considering the damping of the OMD modes; the least damped mode corresponds to the shedding mode of the cylinder at the frequency of interest. 
Since the modes are complex, an instantaneous phase value, $ \phi_n(\textbf{t}) $, can be calculated for a particular mode. 
The angle of the complex coefficients corresponding to the shedding mode of the cylinder, $ \angle{c_{shedding}} $, acts as the instantaneous phase signal for the triple decomposition of the velocity field.

\section{Near-wake coherence suppression using a splitter plate} \label{sec:split}

\subsection{Splitter plate design}

\begin{figure}
	\hspace{-0.5cm}
	\centering
	\begin{tikzpicture}
	\node[anchor = west, inner sep=0] (0) at (3,1.27)
	{\includegraphics[width = 0.6\textwidth]{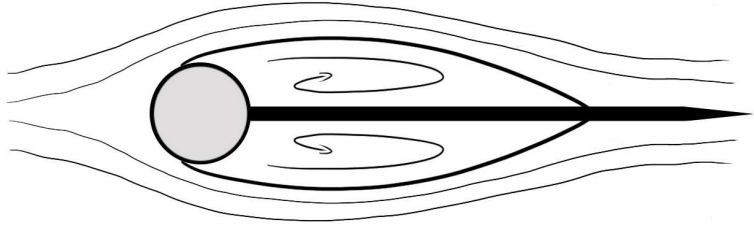}};
	\end{tikzpicture}
	\caption{Schematic of a splitter plate satisfying the ``infinitely'' long condition.}
	\label{fig: splitter plate cartoon}
\end{figure}

The suppression of coherence was achieved through the use of an ``infinitely long'' splitter plate that was attached to the rear of the cylinder.
The ``infinitely long'' condition is said to be achieved when the reattachment point is on the splitter plate itself, since a further increase in length of the splitter plate has no effect on the flow pattern.
This is illustrated through a flow schematic in figure \ref{fig: splitter plate cartoon}.
\cite{Apelt1975} show that this condition occurs when $ L/d > 5 $ for a circular cylinder. Where, $ L $, is the length of the splitter plate and $ d $, is the diameter of the circular cylinder.
They also mention that for an ``infinitely long'' splitter plate, vortices that form over the plate are small and random, no coupling between the two sides of the splitter plate was witnessed, and that the shed wake was narrower and more steady.
By installing such a splitter plate, regular vortex shedding from the cylinder was eliminated.
However, \cite{Apelt1975} noticed a well developed vortex street formed far downstream ($ \sim 17d $) even with a long splitter plate such as this. 
They associated this behaviour to the cylinder/splitter plate combination giving rise to its own unique vortex street.

%Due to the need for the splitter plate to house a 2.39 mm hypodermic tube to transport fluorescent dye in an initial design, the thickness, $ t $, of the splitter plate had to be increased to 5 mm. 
Investigators in the past have tended to stick to a thickness to diameter ratio, $ t/d \lesssim 0.1 $ \citep{Apelt1975,Bearman1965,Roshko1961}. 
Due to experimental constraints the thickness in the current experiment, $ t/d \approx 0.17 $, meant that a streamlining section was added to the trailing edge of the splitter plate to prevent vortex shedding arising from a blunt trailing edge.
The diffuser angle for the streamlining section was designed to equal $ 3^{\circ} $ to prevent any flow separation. 
The complete length of the splitter plate was therefore equal to, $ L/d = 6.59 $.
The Rhodamine 6G dye was still released from the rear face of the cylinder but two pumps connected to two separate release holes were now used.
This allowed the dye to be released into both of the recirculation zones on either side of the splitter plate and hence allowed the dye to fully mix and cover the complete extent of the wake.

\subsection{Coherence suppression}

\begin{figure}
	\hspace{-0.5cm}
	\begin{tikzpicture}
	\node[anchor= west,inner sep=0] (0) at (0,1.27)
	{\includegraphics[width = 0.5\textwidth]{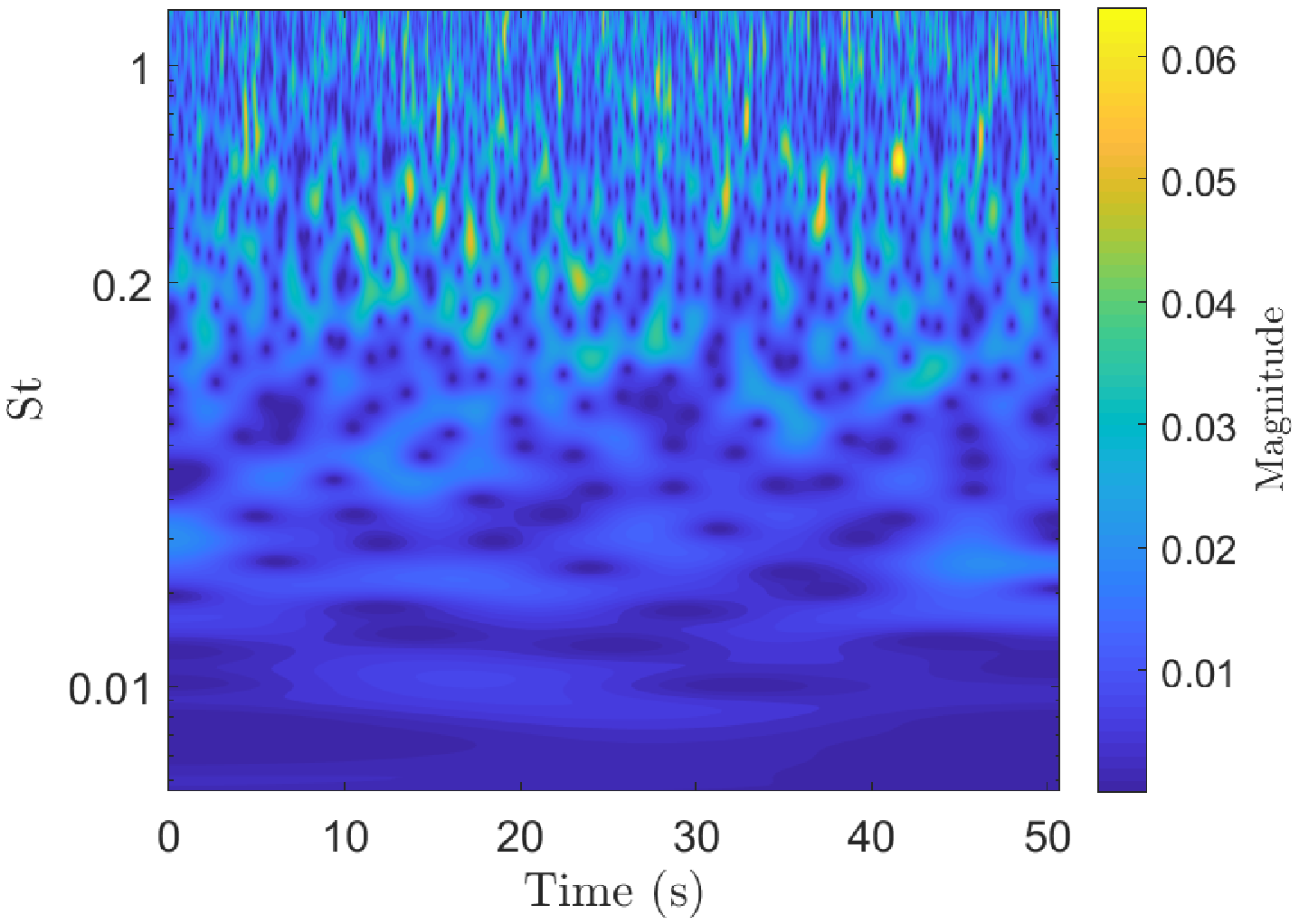}
		\includegraphics[width=0.5\textwidth]{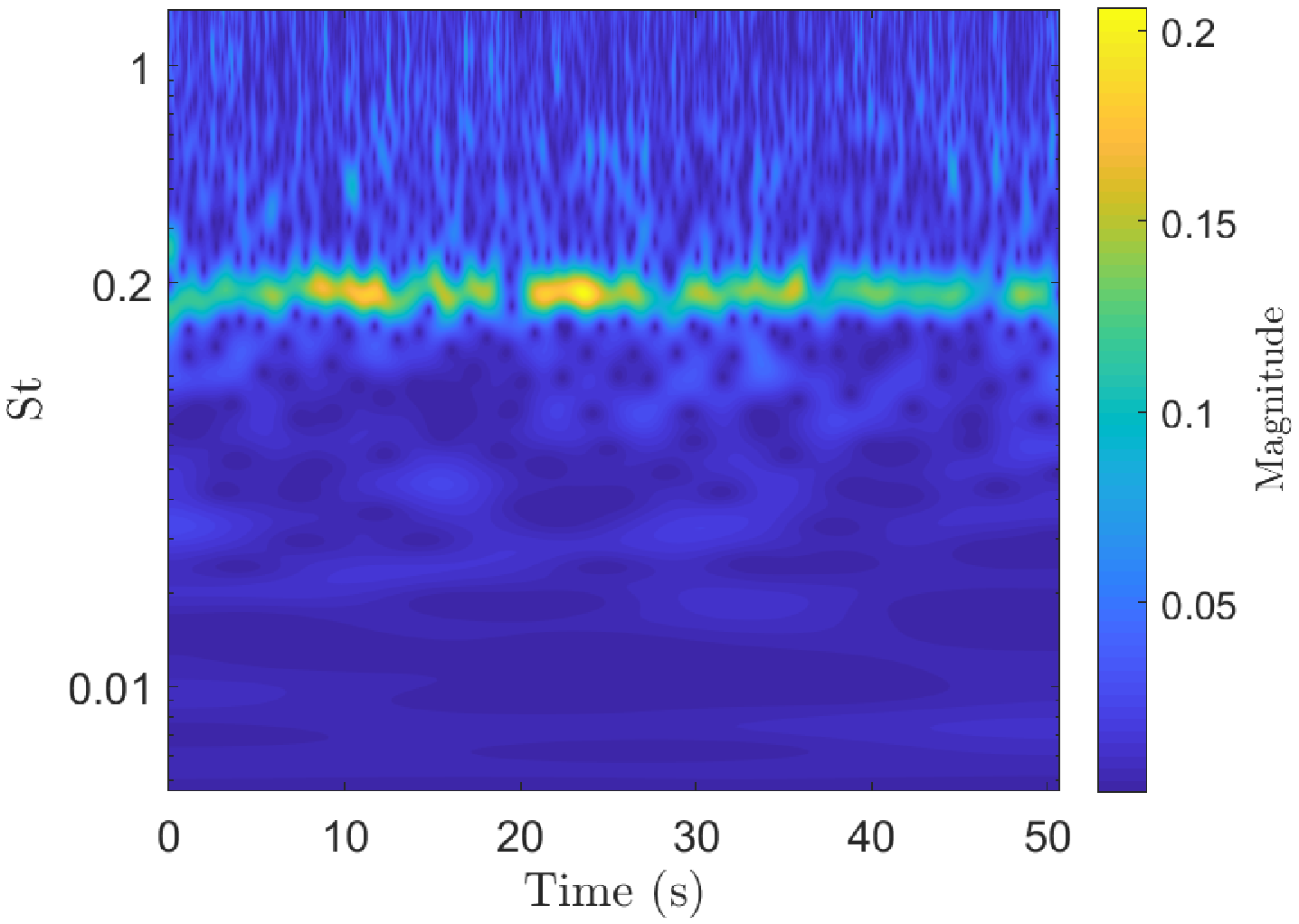}};
		\node[anchor= west,inner sep=0] (1) at (4.5,-6)
		{\includegraphics[width=0.5\textwidth]{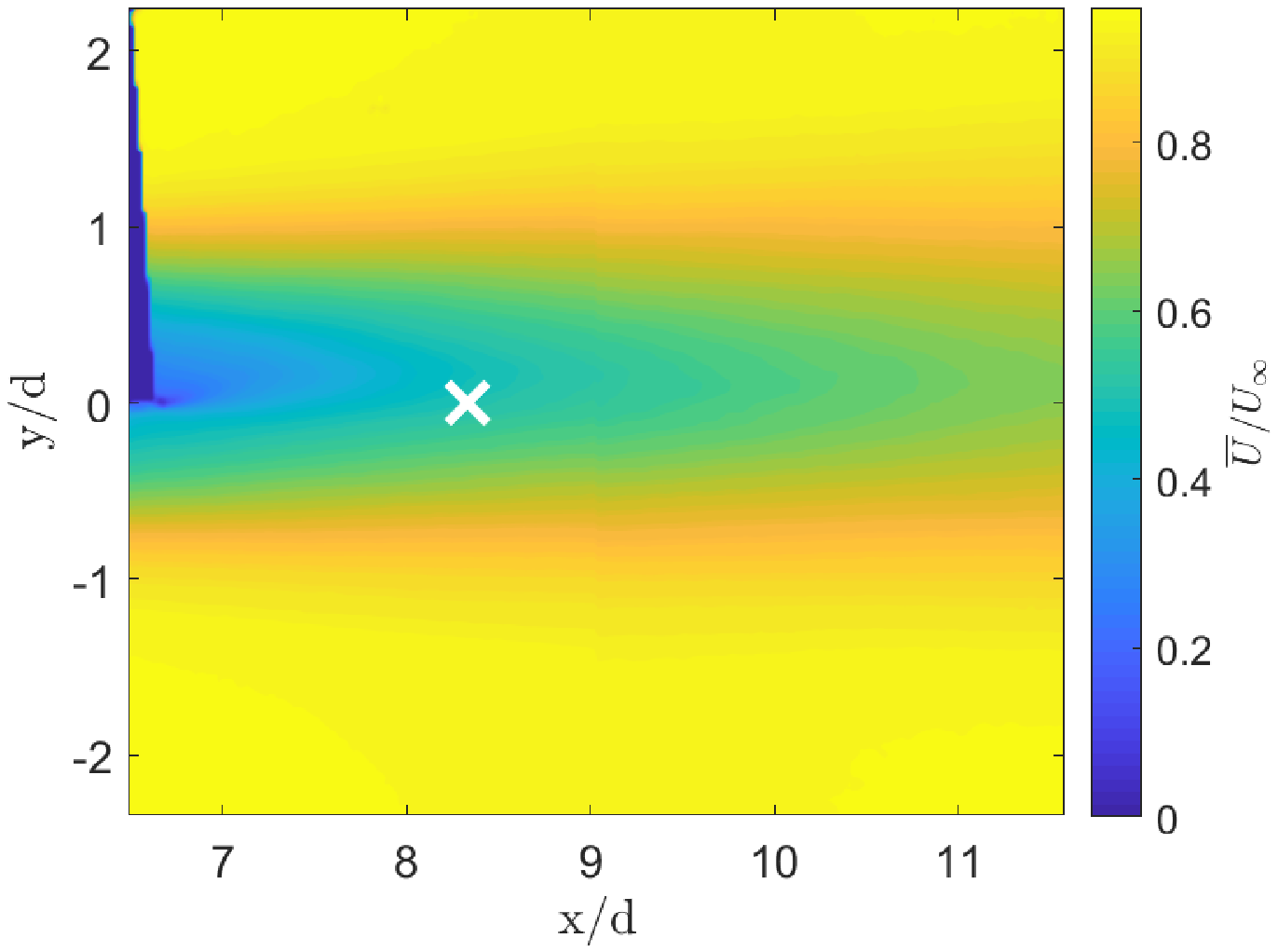}};
	\node (a) at (4.2,-2.35) {(\textit{a}) With splitter plate};
	\node (b) at (12.7,-2.35) {(\textit{b}) Without splitter plate};
	\node (c) at (8.45,-9.7) {(\textit{c}) Velocity probe location};
	\end{tikzpicture}
	\caption{(\textit{a}) Magnitude scalogram of the wavelet transform of $v'$ at the probe location taken during a run with the splitter plate installed. (\textit{b}) Represents the same scalogram, although this time for a clean cylinder without a splitter plate. (\textit{c}) The white cross defines the location of the probe where the spectral signature is analysed. The background field is the mean velocity in the streamwise direction for the run with the splitter plate installed.}
	\label{fig: Coherence supp wavelet}
\end{figure}

To determine whether the coherent structures have been fully eliminated the spectral signature of the wake was extracted and analysed downstream of the trailing edge of the splitter plate.
Figures \ref{fig: Coherence supp wavelet}(\textit{a}) and \ref{fig: Coherence supp wavelet}(\textit{b}) depict the frequency-time domain magnitude scalogram of the continuous wavelet transform of the transverse fluctuating velocity, $v'$, for runs with and without the splitter plate installed respectively. Note that the transform was conducted using a Morse wavelet.
The velocity is probed at the location defined by the white cross in figure \ref{fig: Coherence supp wavelet}(\textit{c}).
It is clear to see that no clear shedding band is visible in the magnitude scalogram across the whole time series in the wake of the splitter plate (see \ref{fig: Coherence supp wavelet}(\textit{a})).
This result is in agreement with the results published by \cite{Roshko1961}, who comment on the fact that there is no peak in the spectrum when a splitter plate is installed.
This is in complete contrast to the spectral signature when no splitter plate is installed.
As expected, a strong shedding band is visible in figure \ref{fig: Coherence supp wavelet}(\textit{a}) for the entire duration of the run.
This is a clear indication that the splitter plate was successful in suppressing coherent structures in the near-field of the cylinder-splitter wake.

\vspace{0.5cm}
\noindent
\textbf{Acknowledgements}: KK would like to acknowledge financial support given by the Engineering and Physical Sciences Research Council through Grant No. EP/R512540/1.
OB would like to acknowledge funding through an Engineering and Physical Sciences Research Council Fellowship, Grant No. EP/V006436/1.

%\bibliographystyle{plainnat}
%\bibliography{NearWakeBib}% Produces the bibliography via BibTeX.

\end{document}